\documentclass[aps,prd,preprint,superscriptaddress,showpacs,showkeys,nofootinbib]{revtex4-1}

\usepackage{hyperref}
\usepackage[latin1]{inputenc}
\usepackage[T1]{fontenc}
\usepackage[english]{babel} 
\usepackage{amsfonts,amssymb,amsmath,amscd,mathtools,slashed}
\usepackage[amsthm,thmmarks,amsmath]{ntheorem}
\usepackage{enumerate}
\usepackage{color}
\usepackage{verbatim}
\usepackage{graphicx}

\usepackage{soul}

\usepackage{relsize}

\theoremstyle{definition}

\newcommand{\norm}[1]{\left\Vert {#1} \right\Vert}
\newcommand{\abs}[1]{\left\vert {#1} \right\vert}

\newcommand{\sN}{{\mathbb N}}              
\newcommand{\sZ}{{\mathbb Z}}
\newcommand{\sR}{{\mathbb R}}
\newcommand{\sC}{{\mathbb C}}

\newcommand{\A}{\mathcal{A}}

\newcommand{\D}{\mathcal{D}}

\renewcommand{\H}{\mathcal{H}}

\newcommand{\N}{\mathbb{N}}
\renewcommand{\P}{\mathcal{P}}

\newcommand{\R}{\mathcal{R}}

\DeclareMathOperator{\Ker}{Ker}

\DeclareMathOperator{\Tr}{Tr}

\DeclareMathOperator{\Vol}{vol}
\DeclareMathOperator{\ord}{ord}
\DeclareMathOperator{\Sd}{Sd}
\DeclareMathOperator{\Res}{Res}
\newcommand{\ahd}{(\mathcal{A,H,D})}            

\newcommand{\DD}{\mathcal{D}}           
\newcommand{\Dslash}{{\DD \mkern-11.5mu/\,}} 

\newcommand{\vc}{\vcentcolon =}

\newcommand{\bra}[1]{\langle #1|}                  
\newcommand{\ket}[1]{| #1 \rangle}                  

\newcommand{\OO}{\mathcal{O}}   

\newcommand{\tzero}{\; \underset{\sigma \downarrow 0}{\sim} \;}      

\newcommand{\simp}{^\text{sm}}
\newcommand{\spin}{^\text{sp}}
\newcommand{\scl}{^\text{sc}}

\newcommand{\Psim}{\P_q\simp}
\newcommand{\Pspn}{\P_q\spin}
\newcommand{\Pscl}{\P_q\scl}

\newcommand{\h}{\mathfrak{h}}

\renewcommand{\Re}{\mathrm{Re}}

\newcounter{mnotecount}[section]
\renewcommand{\themnotecount}{\thesection.\arabic{mnotecount}}
\newcommand{\mnote}[1]
{\protect{\stepcounter{mnotecount}}$^{\mbox{\footnotesize
$
\bullet$\themnotecount}}$ \marginpar{
\raggedright\tiny\em
$\!\!\!\!\!\!\,\bullet$\themnotecount: #1} }

\definecolor{darkgreen}{rgb}{0,.5,0}

\begin{document}


\title{Spectral dimensions and dimension spectra of \break quantum spacetimes}

\author{Micha{\l} Eckstein}
\email{michal@eckstein.pl}
\affiliation{Institute of Theoretical Physics and Astrophysics,
National Quantum Information Centre, Faculty of Mathematics, Physics and Informatics,
University of Gda\'nsk, ul. Wita Stwosza 57, 80-308 Gda\'nsk, Poland}
\affiliation{Copernicus Center for Interdisciplinary Studies, ul. Szczepa\'nska 1/5, 31-011 Krak\'ow, Poland}

\author{Tomasz Trze\'sniewski}
\email{t.trzesniewski@uj.edu.pl}
\affiliation{Institute of Theoretical Physics, Jagiellonian University, ul. S. {\L}ojasiewicza 11, 30-348 Krak\'ow, Poland}
\affiliation{Copernicus Center for Interdisciplinary Studies, ul. Szczepa\'nska 1/5, 31-011 Krak\'ow, Poland}

\date{\today}

\begin{abstract}
Different approaches to quantum gravity generally predict that the dimension of spacetime at the fundamental level is not 4. The principal tool to measure how the dimension changes between the IR and UV scales of the theory is the spectral dimension. On the other hand, the noncommutative-geometric perspective suggests that quantum spacetimes ought to be characterised by a discrete complex set -- the dimension spectrum. Here we show that these two notions complement each other and the dimension spectrum is very useful in unravelling the UV behaviour of the spectral dimension. We perform an extended analysis highlighting the trouble spots and illustrate the general results with two concrete examples: the quantum sphere and the $\kappa$-Minkowski spacetime, for a few different Laplacians. In particular, we find out that the spectral dimensions of the former exhibit log-periodic oscillations, the amplitude of which decays rapidly as the deformation parameter tends to the classical value. In contrast, no such oscillations occur for either of the three considered Laplacians on the $\kappa$-Minkowski spacetime.
\end{abstract}



\maketitle

\section{Introduction}

The concept of spacetime, understood as a differentiable manifold, has proven to be extremely fruitful in modelling gravitational phenomena. However, it is generally expected that the smooth geometry breaks down at small scales or high energies, due to the quantum effects. Consequently, many of the familiar notions, such as causality, distance or dimension, have to be refined within the adopted new mathematical structure.

An essential property of the hypothetical quantum theory of gravity, as well as a useful input for constructing it, is the ability to provide meaningful and testable predictions concerning deviations of physics from general relativity. The first step in this direction can be done by characterizing the structure of (static) quantum spacetime, which replaces the classical differentiable manifold, but can be seen as a certain tangible generalisation of the latter. This is possible only if an unambiguous notion of a spacetime could be provided. In some of the approaches, such a notion is preserved only at the intermediate --- semiclassical --- level, while in the full theory spacetime breaks down into discrete elements, determined either by the fundamental length scale or a regularization cutoff (see \cite{Oriti:2018ly} for a conceptual discussion). In analogy to systems in condensed matter physics, configurations of the underlying ``atoms of spacetime'' may form different phases, while (classical) continuous spacetime should emerge in the limit in at least one of them. Other phases will naturally share some features with the classical phase. Let us stress in this context the distinction between the continuum limit, which is a transition from a discrete proto-spacetime to the continuous (but still quantum) spacetime, and the classical limit, in which we completely recover familiar manifolds of general relativity.

In the recent years, calculations of the effective number of spacetime dimensions have become an ubiquitous method to characterise the quantum spacetime. This is one of only a few tools allowing us to find some order in the diverse landscape of quantum gravity models \cite{Mielczarek:2018ty}, whose predictions are notoriously difficult to compare. The dimension can be defined in many different ways, some of which are based on mathematical assumptions and some on physical concepts, such as thermodynamics \cite{Carlip:2017dy}. The most popular notion remains the so-called \emph{spectral dimension}, which can be seen as determined by the mathematical properties of spectral geometry or by a physical (fictitious) diffusion process. The advantage of the spectral dimension in the latter context is its dependence on a parameter (auxiliary diffusion time) that can be identified with the length scale at which the geometry is probed. Therefore, it is naturally interpreted as a measure of how the dimension of spacetime changes with scale.

Starting with the seminal paper \cite{Ambjorn:2005se} (updated in \cite{Coumbe:2015es}), belonging to causal dynamical triangulations approach to quantum gravity, the spectral dimension has been calculated for Ho\v{r}ava-Lifshitz gravity \cite{Horava:2009st}, asymptotic safety scenario \cite{Lauscher:2005fy}, nonlocal quantum gravity \cite{Modesto:2012sy}, spin foam models \cite{Steinhaus:2018el} (and kinematical states of loop quantum gravity in general \cite{Calcagni:2015ds}), causal sets \cite{Carlip:2015dy,Belenchia:2016ss} and multifractional spacetimes \cite{Calcagni:2012dy}. The almost universal prediction is the dimensional reduction at the smallest scales (the UV limit) to the value of $2$, for which general relativity would actually become power-counting renormalizable. On the other hand, in some cases values different from $2$ are obtained in the UV limit, especially in non-classical phases of models with non-trivial phase diagram. The situation is similar for quantum gravity in less than $3+1$ topological dimensions \cite{Benedetti:2009se,Horava:2009st,Lauscher:2005fy,Calcagni:2012dy} (see also \cite{Alesci:2012ay}). 

Quantum spacetime often turns out to be described in terms of broadly understood noncommutative geometry, which is also sometimes treated as a stand-alone approach to quantum gravity. A particular example is the $\kappa$-Minkowski spacetime \cite{Majid:1994by}, associated with $\kappa$-Poincar\'{e} algebra \cite{Lukierski:1991qa,Lukierski:1992ny} and widely considered in doubly/deformed special relativity and relative locality. As it was shown in \cite{Arzano:2014de}, the small-scale behaviour of the spectral dimension of $\kappa$-Minkowski spacetime (first calculated in \cite{Benedetti:2009fe}) depends on the Laplacian, which can be chosen according to several distinct principles. For $3+1$-dimensional $\kappa$-Minkowski spacetime, there are at least three possibilities in the UV: the dimension decreasing to $3$, growing to $6$ or diverging. However, as we will discuss in Subsec.~\ref{sec:4.4}, there may be a way to reconcile these contrasting results. Let us also note that \cite{Arzano:2017ua} presents an example of a noncommutative toy model (with ${\rm U}(1) \times {\rm SU}(2)$ momentum space) that exhibits the dimensional reduction to $2$, i.e. the value obtained in the approaches to quantum gravity mentioned before. 

Recently, it has also been suggested \cite{CalcagniMulti,Calcagni:2017cy}, in the context of multifractional theories \cite{Calcagni2017review}, that the dimension of quantum spacetimes can acquire complex values. These, on the other hand, result in log-periodic oscillations in various physical quantities \cite{dunne2012} and can possibly affect the CMB spectrum \cite{Calcagni:2017cy}, introduce a stochastic noise to gravitational waves \cite{Amelino2017} or modify the thermodynamics of photons \cite{DunnePRL}. More generally, complex dimensions (or complex critical exponents) and the corresponding log-periodic oscillations can also arise in the systems with discrete scale invariance, which is observed in many contexts, including some particular cases of holography \cite{Flory:2018dd}, as well as condensed matter physics, earthquakes and financial markets, see e.g. \cite{Sornette:1998ds}.

Meanwhile, it has already been recognised by Connes and Moscovici in 1995 \cite{ConnesMoscovici} that quantum spaces --- understood as spaces determined by noncommutative algebras of observables --- ought to be characterised by a discrete subset of the complex plane -- the \emph{dimension spectrum}, rather than a single number. More precisely, in noncommutative geometry the dimension spectrum is defined as the set of poles of the spectral zeta functions of geometrical provenance. These, on the other hand, are intimately connected with the celebrated heat trace expansion via the Mellin functional transform (cf. \cite{Gilkey1,Gilkey2} and also \cite{BookSA}). Both spectral zeta functions and heat trace expansions are indispensable tools in quantum field theory \cite{ZetaQFT,AnalyticQFT,HeatQFT}, also in its noncommutative version \cite{Wulkenhaar}. The heat trace can be utilised to compute the one-loop effective action and allows to study the short-distance behaviour of propagators, along with quantum anomalies and some non-perturbative effects \cite{VassilevichReport}. Consequently, it is justified to expect that the structure of the entire dimension spectrum of a given noncommutative geometry is relevant for physics.

Among the noncommutative spaces with known dimension spectra an interesting example is provided by the Podle\'s quantum sphere \cite{Podles}. For a particular choice of geometry \cite{dab_sit}, the dimension spectrum turns out to have surprising features \cite{PodlesSA}: Firstly, it exhibits the dimension drop (also called the dimensional reduction) from 2 to 0. Secondly, the associated spectral zeta-function contains poles outside the real axis, suggesting self-similarity. Thirdly, these poles are of second order, which is characteristic for spaces with conical singularities \cite{Lescure}. Finally, the corresponding heat trace expansion turns out to be convergent -- in sharp contrast to the case of smooth manifolds, where it is only asymptotic. As we will show, the above features are shared by two other geometries (i.e. other Laplacians) on the quantum sphere but the fine details of their heat trace expansions are different. The latter fact strengthens the observation made in \cite{Arzano:2014de} by one of us in the case of $\kappa$-Minkowski space that the spectral dimension characterises a given quantum space equipped with a specific Laplacian.

Let us note that the effective dimensionality of spaces with the topological dimension lower than 4 is relevant not only from the perspective of toy models of lower dimensional (quantum) gravity but also in the context of the problem of entanglement entropy \cite{Arzano:2017fy}. The reason is that the latter can be derived from the heat trace over the boundary of some region of space \cite{Nesterov:2011gy}.

The purpose of this work is to revisit the concept of the spectral dimension from the perspective of the dimension spectrum. We show that the latter is a valuable rigorous tool to study the UV behaviour of the spectral dimension. To this end, we firstly provide, in Section \ref{sec:gen}, the definitions of both concepts and highlight the trouble spots. Then, in Section \ref{sec:QS}, we compute the dimension spectra and spectral dimensions of three Laplacians on the Podle\'s quantum sphere \cite{Podles}. We show that the dimension drop observed by Benedetti \cite{Benedetti:2009fe} has a finer structure with the square-logarithmic decay and log-periodic oscillations. For values of the deformation parameter $q$ close to the classical value 1 the amplitude of these oscillations becomes very small and they are invisible in numerical plots. On the other hand, their presence is clearly attested for any $q$ by non-real numbers in the dimension spectrum. Next, in Section \ref{sec:kMink}, we study the dimension spectra of three different Laplacians on the $\kappa$-Minkowski spacetime in 2, 3 and 4 topological dimensions. We utilise these to identify the leading and subleading short-scale behaviour of the spectral dimensions obtained in \cite{Arzano:2014de}. This example uncovers an ambiguity in the definition of the spectral dimension related to the order of the `Laplacian-like' operator. We summarise our findings in Section \ref{sec:discussion} and discuss their consequences for model-building in quantum gravity.

\section{\label{sec:gen} Two faces of dimensionality}

\subsection{Spectral dimension of a diffusion process}

The usual perspective in quantum gravity is to introduce the spectral dimension as a characteristic of the fictitious diffusion (or random walk) process on a given configuration space. Let us first consider a diffusion process on a Riemannian manifold $(M,g)$ of topological dimension $d$, which is described by the heat equation
\begin{align}\label{eq:3.1}
\frac{\partial}{\partial\sigma} K(x,x_0;\sigma) + \Delta K(x,x_0;\sigma) = 0\,,
\end{align}
with a second order differential operator $\Delta$ in variable $x$ and an auxiliary time variable $\sigma \geq 0$ (playing the role of a scale parameter). In general, the operator $\Delta$ does not need to be the standard Laplacian $-g^{\mu\nu} \nabla_\mu \nabla_\nu$, $\mu,\nu = 1,\ldots,d$ --- it can be a Laplace-type operator \cite{VassilevichReport} or even a pseudodifferential one (cf. \cite{Shubin}).

In order to solve Eq. \eqref{eq:3.1} one needs to impose appropriate initial/boundary conditions. Typically, one chooses the initial condition of the form
\begin{align}\label{eq:3.2}
K(x,x_0; \sigma = 0) = \frac{\delta^{(d)}(x - x_0)}{\sqrt{|\det g(x)|}}\,.
\end{align}

In particular, in the case of $d = 4$ and the flat Euclidean metric, the solution to (\ref{eq:3.1}) can be expressed as a (inverse) Fourier transform
\begin{align}\label{eq:3.3}
K(x,x_0;\sigma) = \int \frac{d^4 p}{(2\pi)^4}\, e^{i p_\mu (x - x_0)^\mu} e^{-\sigma {\cal L}(p)}\,,
\end{align}
where ${\cal L}$ is the momentum space representation of $\Delta$.

To characterise the diffusion process (\ref{eq:3.1}) we may use the return probability
\begin{align}\label{eq:3.4}
{\cal P}(\sigma) = \frac{1}{\mathrm{vol}\, V} \int_V  d^4x \sqrt{|\det g|}\, K(x,x;\sigma)\,, \qquad \sigma > 0\,,
\end{align}
where we integrate over a fiducial volume $V$. (It factorises in the leading term and therefore can be taken to infinity if needed). ${\cal P}(\sigma)$ is the probability that after the time $\sigma$ the diffusion will return to the same point $x \in V \subset M$. The spectral dimension is now taken to be a function of the scale parameter $\sigma$ defined as
\begin{align}\label{eq:3.5}
d_S(\sigma) \vc -2 \, \frac{\partial\log{\cal P}(\sigma)}{\partial\log\sigma} = -\frac{2\sigma}{{\cal P}(\sigma)}\, \frac{\partial{\cal P}(\sigma)}{\partial\sigma}\,.
\end{align}

In the case of $M$ being a flat Euclidean space of topological dimension $d$, we have $d_S(\sigma) = d$ for all $\sigma$. Therefore, in general, if $d_S(\sigma) \in \sN$ for a given value of $\sigma$, it can be interpreted as the effective dimension such that the ordinary diffusion process in $d_S(\sigma)$-dimensional Euclidean space would approximately behave as the $\Delta$-governed diffusion on $M$. If we choose the appropriate Laplacian, small values of $\sigma$ allow us to probe the ultraviolet structure of $M$, while large ones correspond to its infrared geometry. However, for sufficiently large $\sigma$ the function (\ref{eq:3.5}) becomes sensitive to the finite size of $M$ and the curvature of $g$. 

The original definition \eqref{eq:3.5} applies solely in the context of Riemannian manifolds. The departure from smooth geometry requires a suitable generalisation.

Within the framework of deformed relativistic symmetries, the noncommutative geometry of spacetime is accompanied by a curved momentum space. Typically, the latter is a non-Abelian Lie group, equipped with an invariant Haar measure $\mu$. This suggests a natural generalisation of formula \eqref{eq:3.3} to (cf. \cite{Alesci:2012ay,Arzano:2014de,Arzano:2017ua})
\begin{align}\label{eq:3.6}
K(x,x_0;\sigma) = \int \frac{d\mu(p)}{(2\pi)^4}\, e^{i p_\mu (x - x_0)^\mu} e^{-\sigma {\cal L}(p)}\,,
\end{align} 
representing the noncommutative Fourier transform (i.e. inverse of the group Fourier transform) of the function $e^{-\sigma {\cal L}}$.

More generally, one can adopt the definition of a heat operator $e^{-\sigma T}$, which applies for any closed, possibly unbounded, operator $T$ acting on a separable Hilbert space $\H$ (cf. \cite{Shubin} and \cite[Appendix A]{BookSA}). If $T$ is bounded from below and $e^{-\sigma T}$ is trace-class, one defines the \emph{heat trace} (or the `return probability') of an abstract operator $T$ as
\begin{align}\label{heat_gen}
\P(\sigma) \vc \Tr_{\H} e^{-\sigma T} = \sum_{n=0}^{\infty} e^{-\sigma \lambda_n(T)},
\end{align}
where $\lambda_n(T)$ are the eigenvalues of $T$ counted with their multiplicities.

For a compact Riemannian manifold $M$ and $T = \Delta$, the trace can be computed via the standard integral kernel methods and one obtains (from now on we drop the normalization)
\begin{align*}
\Tr_{L^2} e^{-\sigma \Delta} = \int_M d^4x \sqrt{|\det g|}\, K(x,x;\sigma) = {\cal P}(\sigma)\,,
\end{align*}
for any $\sigma > 0$.

Formula \eqref{heat_gen} allows us to extend the notion of the spectral dimension \eqref{eq:3.5} beyond the realm of smooth manifolds. For an abstract operator $T$ one has
\begin{align}\label{dS_eig}
d_S(\sigma) = 2\sigma\, \frac{\Tr_{\H} T e^{-\sigma T}}{\Tr_{\H} e^{-\sigma T}} = 2\sigma\, \frac{\sum_{n=0}^{\infty} \lambda_n(T)\, e^{-\sigma \lambda_n(T)}}{\sum_{n=0}^{\infty} e^{-\sigma \lambda_n(T)}}\,.
\end{align}

Let us now point a few trouble spots with usage of \eqref{heat_gen} and hence the spectral dimension:

\begin{enumerate}

	\item Firstly, one needs to make sure that formula \eqref{heat_gen} is well defined. The trace-class property of $e^{-\sigma T}$ for all $\sigma > 0$ is guaranteed on general grounds if $T$ is a classical pseudodifferential operator on a compact manifold $M$ \cite{Gilkey1,Gilkey2}, but it may fail, for instance, on infinite dimensional spaces \cite{ConnesMarcolli}.
	
	\item If the spacetime manifold $M$ is not compact, then $e^{-\sigma T}$ is typically not trace-class (actually, not even compact) even if $T$ is an honest classical pseudodifferential operator. Consequently, to define the heat trace one needs an IR cut-off in the form of a trace-class operator $F$
	\begin{align}\label{heat_gen_F}
	\P(\sigma,F) \vc \Tr_{\H} F \, e^{-\sigma T}.
	\end{align}
	On a manifold, one can simply take $F$ to be a function projecting on a compact fiducial volume $V$, as done in formula \eqref{eq:3.4}. Then, after restoring the normalisation, $V$ can eventually be taken to infinity, showing the independence of the spectral dimension on the IR regularisation. On the other hand, there is no reason to assume that a similar factorisation would take place outside of the realm of manifolds. Although it can be demonstrated for specific examples, such as the $\kappa$-Minkowski spacetime which we consider in Sec. \ref{sec:kMink}, in general one should expect to encounter the notorious IR/UV mixing problem \cite{IRUV}.
	
	\item The multiplicative factor 2 in Eq. \eqref{eq:3.5} originates from the fact that the Laplacian is a second order differential operator. This can be easily adapted if $T$ is (pseudo)differential operator of any order $\eta > 0$ by redefining
	\begin{align}\label{dSm}
    d_S(\sigma) \vc -\eta \frac{\partial\log{\cal P}(\sigma)}{\partial\log\sigma}.
    \end{align}
	However, beyond the safe realm of smooth manifolds, the order of $T$ is not a priori defined and \eqref{dSm} becomes ambiguous. We shall illustrate this problem in Section \ref{sec:kMink}.
	
	\item The direct computation of the spectral dimension from formula \eqref{dS_eig} requires full knowledge about the spectrum of $T$, which is seldom granted. One could resort to asymptotic formulae for heat traces (cf. Formula \eqref{heat_pdo}) to unveil the small-$\sigma$ behaviour of the spectral dimension, but the result can be very misleading if one quits the UV sector \cite{ET_NEW}.
	
	\item A consistent interpretation of $d_S(\sigma)$ as a scale-dependent dimension of spacetime requires it to reach the ``classical value'' in the IR sector. However, the latter is equal to 4 only in the very specific instance of $\sR^4$, in which case actually $d_S(\sigma) = 4$ independently of the value of $\sigma$. If the classical spacetime has compact topology or non-trivial curvature, then either $d_S(\sigma)$ tends to 0 or grows to infinity as $\sigma \to \infty$, depending on whether the operator $T$ has a trivial kernel or not (cf. $\Delta^{\text{sc}}$ versus $\Delta^{\text{sp}}$ on Fig. \ref{dS_spheres}). As for the latter, one can use the notion of the spectral variance \cite{Lisa:2019} to remove the zero mode. In either case, in order to recover the correct dimension in the IR, one has to match the large-scale behaviour of $d_S(\sigma)$ of the quantum model with the corresponding classical spacetime manifold, as it was done for the CDT in \cite{Benedetti:2009se}.
	
	\item Finally, in order to study the spectral dimension of spacetime, which is characterized by a Lorentzian metric, one first has to perform the Wick rotation of it, i.e. an analytic continuation to the Euclidean signature. This is a rather cumbersome procedure even in the case of curved pseudo-Riemannian manifolds and it is likely to be even more problematic in quantum spacetime models (see, however, \cite{LizziWick}). We shall not explore this issue here since the considered examples are either Euclidean from the start (quantum spheres) or have the well defined Euclidean counterparts ($\kappa$-Minkowski momentum spaces \cite{Arzano:2014de}).
	
\end{enumerate}

\subsection{Dimension spectrum from asymptotic expansion}

Let us now restart with the spacetime modelled by a Riemannian manifold $M$ and turn towards the notion of the dimension spectrum.

Abundant information about the geometry of $M$ can be learned from the celebrated heat kernel expansion \cite{Gilkey1,Gilkey2,VassilevichReport}:
\begin{align}\label{heat_pdo}
\P(\sigma) \tzero \sum_{k=0}^\infty a_k(T)\, \sigma^{(k-d)/\eta} + \sum_{\ell=0}^{\infty} b_\ell(T)\, \sigma^{\ell}\, \log\sigma\,,
\end{align}
where $\eta$ is the order of the pseudodifferential operator $T$.
A few comments about formula \eqref{heat_pdo} are in order:

Firstly, the infinite series in formula \eqref{heat_pdo} are asymptotic series, which are in general \emph{divergent for any $\sigma > 0$}. Nevertheless, the formula has a precise meaning as an asymptotic expansion (cf. \cite{Elizaldebook} and \cite[Section 2.5]{BookSA}). It provides accurate information about the small-$\sigma$ asymptotic behaviour of $\P(\sigma)$, 
but in general it fails for larger values of $\sigma$ \cite{ET_NEW}.

Secondly, if $T$ is a differential operator, the coefficients $b_{\ell}(T) = 0$ and $a_k(T)$ are locally computable quantities of geometrical origin --- that is they can be expressed as $a_k(T) = \int_{M} \alpha_k^T(x)$. On the other hand, if $T$ is only pseudodifferential, then $a_k(T)$ for $(k-d) \in \eta\N$ are \emph{not} locally computable \cite{GilkeyGrubb}. 

Thirdly, one sees that $d = \dim M$ is encoded in Formula \eqref{heat_pdo} as the leading small-$\sigma$ behaviour, while $a_0(T) \propto \Vol(M)$.
If $T$ is a scalar Laplace-type operator and $M$ has no boundary, then the odd coefficients $a_{2n+1}(T)$  vanish. The second coefficient reads $a_2(T) = \tfrac{1}{6} \, (4 \pi)^{-d/2} \int_M d^d x\sqrt{g(x)} \R(x)$, where $\R$ is the scalar curvature, whereas $a_{2n}(T)$ for $n \geq 2$ involve higher order invariants constructed from the Riemann tensor. If $T$ acts on a vector bundle $E$ over $M$, then $a_{n}(T)$ involve also the curvature of $E$ --- see \cite{VassilevichReport} for a complete catalogue.

Finally, Formula \eqref{heat_pdo} extends to the non-compact setting. As mentioned earlier, this requires an IR regularising operator --- typically, a compactly supported smooth function $f$ on $M$. In such a case, $\P(\sigma,f)$ still admits an asymptotic expansion of the form \eqref{heat_pdo} but its coefficients now depend on $f$.

Let us now leave the domain of smooth manifolds and trade the pseudo-differential operator for a positive unbounded operator acting on a separable Hilbert space $\H$. In this context, one expects a more general form of the heat trace expansion (cf. \cite{BookSA}):
\begin{align}\label{HeatGen}
\P(\sigma) = \Tr_{\H} e^{-\sigma T} \tzero \sum_{k=0}^\infty \sum_{m \in \sZ} \sum_{n=0}^{p} a_{z(k,m),n}\, (\log\sigma)^n\, \sigma^{-z(k,m)}\,,
\end{align}
for a (discrete, but possibly infinite) set of complex numbers $z(k,m)$.
We define the \emph{dimension spectrum of the operator $T$} as the collection of exponents (i.e. a set of numbers):
\begin{align}\label{SdT}
\Sd(T) \vc \bigcup_{k,m} z(k,m) \subset \sC\,,
\end{align}
whence the number $(p+1)$, capturing the maximal power of $\log \sigma$ terms in \eqref{HeatGen}, is called the \emph{order} of the dimension spectrum $\ord \Sd(T)$. It is also useful to define the maximal (real) dimension in the spectrum
\begin{align}\label{dSd}
d_{\Sd} \vc \sup_{z \in \Sd}\, {\rm Re}(z)\,.
\end{align}

From Formula \eqref{heat_gen} one immediately reads out that if $T$ is a classical pseudodifferential operator, then $\Sd(T) \subset \tfrac{1}{\eta} (d - \N) = \{(d-k)/\eta\, \vert\, k \in \N\}$ and $\ord\Sd(T) \leq 2$, where $\eta = \eta(T)$ is the order of $T$ and $d$ is the dimension of the underlying manifold. In particular, if $T$ is a differential operator, then $\ord\Sd(T) = 1$. In either case we have $d_{\Sd} = d/\eta$, as expected. 

Dimension spectra of higher order can be found beyond the realm of classical pseudodifferential operators \cite{Lesch,Lescure,Gil2008,BookSA}. In particular, dimension spectra of order 3 were found for Fuchs-type operators on manifolds with conical singularities \cite{Lescure}. Surprisingly enough, the same feature was discovered in the very different context of the quantum sphere \cite{JapanHeat,PodlesSA} (cf. Section \ref{sec:QS}). Meanwhile, the presence of complex numbers in $\Sd$, typical for fractal spaces \cite{Lapidus}, is interpreted as a signature of the self-similar structure \cite{Kellendonk}. In view of formula \eqref{HeatGen} this implies in turn that the heat trace exhibits oscillations \cite{dunne2012,DunnePRL}. 


The above definition of dimension spectrum is borrowed from noncommutative geometry \`a la Connes \cite{ConnesNCG}. The central notion of the latter is a spectral triple $\ahd$ consisting of a noncommutative algebra $\A$ of space(time) observables represented on $\H$ and an unbounded operator $\DD$ acting on $\H$, all tied together with a set of axioms. In this context, one talks about the dimension spectrum of a spectral triple $\Sd\ahd$ \cite{ConnesMoscovici} (cf. also \cite[Sec. 1.4]{BookSA}), which is the union of dimension spectra of a family of operators \footnote{The precise statement is: If $\ahd$ is a spectral triple and $\abs{\DD}$ admits an expansion of the form \eqref{HeatGen}, then $\Sd(\abs{\DD}) \subset \Sd\ahd$ and $\ord \Sd(\abs{\DD}) \leq \ord \Sd\ahd$.}.
These originate from the fluctuations of the bare operator $\DD$. 

On the physical side, considering fluctuated $\DD$ amounts to dressing it with all gauge potentials available for a given spectral triple. Thus, one could say that $\Sd(\DD)$ refers to the `pure gravity' scenario. The internal fluctuations of geometry caused by the gauge fields will in general change the dimension spectrum, including its order. They will not, however, change the maximal dimension $d_{\Sd}$ (see \cite[Proposition 4.11]{BookSA}). Observe that the spectral dimension will also change in presence of other `non-gravitational' fields as the bare Laplacian will get dressed by a potential. 

It is also worth noting (see \cite{BookSA,HeatEZ} for the full story) that if a positive operator $T$ admits an expansion of the form \eqref{HeatGen}, then the associated spectral zeta-function $\zeta_T$ defined as
\begin{align}
\zeta_T(s) \vc \Tr T^{-s}, \; \text{ for } \; {\rm Re}(s) \gg 0
\end{align}
admits a meromorphic extension to the whole complex plane. This enjoyable interplay is revealed with the help of the Mellin transform:
\begin{align}\label{Mell}
\int_{0}^\infty \Tr e^{-\sigma T} \, \sigma^{s-1} \, d\sigma = \Gamma(s)\, \zeta_T(s)\,, \; \text{ for } \; {\rm Re}(s) \gg 0\,.
\end{align}

Formula \eqref{Mell} allows us furthermore to retrieve the complete structure of poles of $\zeta_T$. In particular, the set $\Sd(T)$ coincides with the set of poles of the function $\Gamma \cdot \zeta_T$ and, moreover, 
\begin{align}\label{poles}
\forall \, z \in \Sd \quad \underset{s = z}{\Res}(s - z)^n \Gamma(s)\, \zeta_T(s) = (-1)^n n!\, a_{z,n}\,.
\end{align}
Recall that the Gamma function has simple poles at non-positive integers with \linebreak $\Res_{s=-\ell} \Gamma(s) = (-1)^\ell/\ell!$. In summary, a term of order $(\log \sigma)^{n-1} \sigma^{-z}$ in the small-$\sigma$ expansion of $\P(\sigma)$ corresponds to a pole of $\Gamma \cdot \zeta_T$ of order $n$ at $z \in \sC$.

Let us now discuss the properties and problems of the dimension spectrum, as compared with the spectral dimension:

\begin{enumerate}

	\item On top of the problem of checking whether, for a given operator $T$, Formula \eqref{heat_gen} is well-defined, one needs to prove the existence of an asymptotic expansion of the form \eqref{HeatGen}. This is guaranteed if $T$ is a classical elliptic pseudodifferential operator \cite{GilkeyGrubb} (cf. also \cite[Appendix A]{BookSA}). On the other hand, beyond this realm it is a formidable task and no general results are available (see, however, \cite{BookSA,HeatEZ}).
	
	\item In the non-compact case, the dimension spectrum would suffer from the same problems with the IR/UV mixing. Both the coefficients $a_{z,n}$ and the set $\Sd$, as well as its order, would in general depend upon the choice of the IR regularisation \eqref{heat_gen_F}.
	
	\item As in the case of the spectral dimension, the dimension spectrum depends on the order $\eta$ of the operator at hand. This may result in the ambiguous interpretation of $d_{\Sd}$ as the dimension of the underlying space (cf. Section \ref{sec:kMink}).	
	
	\item Because the exponents $z(k,m)$ are in general complex numbers, the heat trace $\P(\sigma)$ will exhibit oscillations as $\sigma$ tends to 0 and hence so will the spectral dimension $d_S$. This oscillatory behaviour of $d_S$ in the UV may be hard to detect in the numerical plots, as we illustrate in Sec. \ref{sec:QS}. Within the dimension spectrum it is, however, well separated from the leading divergence rate in the UV, which is naturally given by $d_{\Sd}$.	 Moreover, we have $d_S(0) = \eta \, d_{\Sd}$, provided that the limit $d_S(0)$ exists.

	\item The dimension spectrum has no issues with the zero modes of $T$ \footnote{If $\ker T$ is non-trivial, the spectral zeta-function $\zeta_T$ is not well defined, but this can be easily circumvented --- see \cite[Eq. (1.1)]{BookSA}.}. Also, in contrast to the spectral dimension, the curvature does not affect the exponents $z(k,m)$ contained in the spectrum but only the coefficients $a_{z,n}$.
	
	\item As in the case of the spectral dimension, the dimension spectrum is not formally defined for spaces with Lorentzian signature (unless they are Wick-rotated). This is because the relevant operators are wave-operators, which are hyperbolic and not elliptic (cf. \cite[Section 2.3]{VassilevichReport} and \cite[Problem (4) in Chap. 5]{BookSA}). Whereas the formal heat kernel expansion \eqref{heat_pdo} does not exist, an analogue of $a_k$'s --- called the Hadamard coefficients --- can be defined (see \cite{Ginoux07}).

\end{enumerate}

In the next Section, we illustrate the (dis)similarities of the two notions of dimensionality via a careful analysis of two classes of examples.

\section{Quantum sphere \label{sec:QS}}

A quantum sphere was first introduced by Podle\'s \cite{Podles} as a quantum homogeneous space of the deformed group $\mathrm{SU}_q(2)$. As a topological space it is described via the complex $^*$-algebra $\A_q$ generated by $A = A^*,\, B$ and $B^*$ subject to the relations
\begin{align}\label{QS_rel}
AB = q^2 BA\,, && AB^* = q^{-2} B^*A\,, && BB^* = q^{-2} A\, (1 - A)\,, && B^*B = A\, (1 - q^2 A)\,,
\end{align}
for a deformation parameter $0 < q < 1$. In the limit $q \to 1$, one recovers the classical algebra of continuous functions on the unit 2-sphere. This abstract algebra is faithfully represented on a Hilbert space $\H_q$ spanned by orthonormal vectors $\ket{l,m}_\pm$, with $m \in \{-l,-l+1,\ldots,l\}$ and $l \in \sN + \tfrac{1}{2}$, mimicking the chiral spinors on $S^2$ \cite{dab_sit}.

The geometry of quantum spheres has been extensively studied within the framework of spectral triples \cite{EquatorialDimSp,DabrowskiQSpheres,AllPodles,EquatorialPodles,dab_sit,PodlesSA}. Among the known geometries particularly interesting is the one equivariant under the action of the Hopf algebra $\mathcal{U}_q(\mathfrak{su}(2))$ \cite{dab_sit}. Its dimension spectrum and heat trace were computed analytically in \cite{PodlesSA}. These turned out to exhibit a number of surprising features:

\begin{itemize}
\item The maximal dimension in the dimension spectrum $d_{\Sd}$ is equal to 0.

\item The spectrum $\Sd$ is of order 3 and the leading term in the expansion \eqref{HeatGen} is $\log^2 \sigma$.

\item $\Sd$ is a regular lattice on the complex plane, which corresponds to $\log$-periodic oscillations of the heat trace and suggests a self-similar structure of the quantum sphere.

\item The expansion \eqref{HeatGen}, expected to be only asymptotic, is actually convergent for all $\sigma$.
\end{itemize}

The quantum sphere served in \cite{Benedetti:2009fe} as a toy example to illustrate the phenomenon of dimension drop in quantum spacetimes. The operator determining the geometry employed in \cite{Benedetti:2009fe} originates from the Casimir operator on the Hopf algebra $\mathcal{U}_q(\mathfrak{su}(2))$. It could be regarded as a `scalar Laplacian', which differs slightly from the `spinor Laplacian' derived from the `Dirac operator' introduced in \cite{dab_sit} --- see Subsection \ref{subsec:QS_comp}. 

Below we present the computations of both the dimension spectrum and the spectral dimension for the above mentioned two Laplacians on the quantum sphere and a third --- `simplified' --- one. 
The latter allows for explicit analytic computations, while capturing the essential small-$\sigma$ behaviour of heat traces for both scalar and spinor Laplacians.

\subsection{Simplified Laplacian \label{subsec:simpl}}

The simplified Dirac operator $\DD_q^S$ on a quantum sphere was introduced in \cite{PodlesSA}. Together with the algebra $\A_q$ and the Hilbert space $\H_q$, it satisfies all of the axioms of a spectral triple. Its square --- the \emph{simplified Laplacian} --- acts on basis vectors of $\H_q$ as
\begin{align*}
\Delta_q\simp \ket{l,m}_\pm = u\, q^{-(2l+1)} \ket{l,m}_\pm\,,
\end{align*}
with $u = u(q) \vc (q^{-1} - q)^{-2}$. The simple exponential form of eigenvalues implies a self-similarity relation $\Delta_q\simp  \vc (\DD_q^S)^2 = u \big\vert \DD_{q^2}^S \big\vert$. 
Note that $\Delta_q\simp$ has no zero modes and does not have a well-defined classical limit $q \to 1$. 

The heat trace associated with $\Delta_q\simp$ reads:
\begin{align}
\Psim(\sigma) &= \Tr_{\H_q} e^{-\sigma \Delta_q\simp} = \sum_{+,-} \sum_{l \in \N + 1/2} \sum_{m=-l}^l {}_\pm\bra{l,m} e^{-\sigma \Delta_q\simp} \ket{l,m}_\pm \notag\\
&= 2 \sum_{l \in \N + 1/2} (2l+1) \exp\left( -\sigma u\, q^{-(2l+1)} \right) = 4 \sum_{n=1}^{\infty} n \exp\left( -\sigma u\, q^{-2n} \right). \label{Psim}
\end{align}
When $\sigma$ tends to infinity, $\Psim(\sigma)$ decays as $e^{-\sigma u q^{-2}}$. However, the small-$\sigma$ behaviour of $\Psim(\sigma)$ cannot be easily deduced from the doubly exponential series in Formula \eqref{Psim}.

On the other hand, the associated zeta function is a simple geometric series:
\begin{align}\label{zeta0}
\zeta_{\Delta_q\simp}(s) = \Tr_{\H_q} (\Delta_q\simp)^{-s} = 4 u^{-s} \sum_{n=1}^{\infty} n\, q^{2 n s} = 4 u^{-s} q^{2s} (1-q^{2s})^{-2},
\end{align}
which is meromorphic on the entire complex plane. It has double poles located solely on the imaginary axis, for $s = \pi i j / \log q$ with $j \in \sZ$.

In this specific case one can deduce, via (the inverse of) Formula \eqref{Mell}, the explicit `non-perturbative' formula for the heat trace \cite[Theorem 4.13]{PodlesSA}:
\begin{align}\label{heat_q0}
\Psim(\sigma) = \tfrac{1}{4 \log^2q}\, \big[ 2 \log^2(u \sigma) + G\big(\log (u\sigma)\big)\, \log (u\sigma) + F\big(\log (u \sigma)\big) \big] + R_{\rm sm}(u\sigma)\,,
\end{align}
where $F$ and $G$ are periodic bounded smooth functions on $\sR$, defined as
\begin{align*}
G(x) &\vc 4 \gamma - 4 \sum_{j \in \sZ^*} \Gamma(\tfrac{\pi i}{\log q}\, j)\, e^{\pi i j x/ \log q}\,, \\
F(x) &\vc \tfrac{1}{3} ( \pi^2 + 6 \gamma^2 - 4 \log^2 q) + 4
\sum_{j \in \sZ^*}\, \Gamma (-\tfrac{\pi i}{\log q} \,j)\, \psi (\tfrac{\pi i}{\log q}\, j)\, e^{\pi i j x/ \log q} \, ,
\end{align*}
with $\sZ^* = \sZ \setminus \{0\}$ and
\begin{align*}
R_{\rm sm}(x) \vc 4\, \sum_{k=1}^\infty \tfrac{(-1)^k\, q^{2k}}{k! (1 - q^{2k})^2}\, x^k\,.
\end{align*}
The symbol $\gamma$ denotes the Euler--Mascheroni constant and $\psi = \Gamma'/\Gamma$ -- the digamma function. All of the series invoked in the above formulae are absolutely convergent on $\sR$.

From Formula \eqref{heat_q0} we can quickly read out the dimension spectrum (see Fig. \ref{fig:QS_Sd} b)):
\begin{align}\label{SimSpec}
\Sd(\Delta_q\simp) = \tfrac{\pi i}{\log q} \sZ \cup (- \sN) = \{ \tfrac{\pi i}{\log q} k\, |\, k \in \sZ \} \cup \{-n\, |\, n \in \sN\}\,, \quad \text{with} \quad \ord\Sd = 3\,.
\end{align}
It coincides with the set of poles of the function $\Gamma \cdot \zeta_{\Delta_q\simp}$, as expected from general theorems discussed around Formula \eqref{Mell}. The third order pole at $s=0$ yields the leading $\log^2 \sigma$ term, the second order purely imaginary poles of $\zeta_{\Delta_q\simp}$ result in the oscillating behaviour captured by $F$ and $G$, whereas the simple poles of $\Gamma$ at negative integers give rise to the remainder $R_{\text{sm}}$.

Let us emphasize that Formula \eqref{heat_q0} is indeed a genuine equality valid for any $\sigma > 0$. This is in sharp contrast with a typical situation of heat trace expansion on a manifold where one has only an asymptotic formula at one's disposal. We can thus compute explicitly the corresponding spectral dimension:
\begin{align}\label{dSsimp}
d_S^{\,q, \text{sm}}(\sigma) = -2 \, \frac{\big[ G'\big(\log (u \sigma)\big) + 4 \big] \log (u \sigma) + F'\big(\log (u \sigma) \big) + G\big(\log (u \sigma) \big) + u \sigma R_{\rm sm}'(u \sigma) }{2 \log^2 (u \sigma) + G \big(\log (u \sigma) \big) \log (u \sigma) + F\big(\log (u \sigma) \big) + R_{\rm sm}(u \sigma)}.
\end{align}

This function is plotted and analysed in Section \ref{subsec:QS_comp} below.

\subsection{Spinor Laplacian \label{subsec:spin}}

Let us now turn to the spinor Laplacian $\Delta_q\spin$. It arises as the square of the Dirac operator $\DD_q$, introduced in \cite{dab_sit}. The latter is a unique $\mathcal{U}_q(\mathfrak{su}(2))$-equivariant operator, which renders a real spectral triple. The spinor Laplacian acts on basis vectors of $\H_q$ as
\begin{align*}
\Delta_q\spin \ket{l,m}_\pm = u\, \left(q^{-(l+1/2)} - q^{(l+1/2)}\right)^2 \ket{l,m}_\pm\,,
\end{align*}
with $u = (q^{-1} - q)^{-2}$, as previously. The operator $\Delta_q\spin$ also does not have a zero mode. In the limit $q \to 1$ it tends (strongly) to the spinor Laplacian on $S^2$ (cf. \eqref{S2spin}). 

The spectral zeta function associated with $\Delta_q\spin$ can easily be computed (cf. \cite[Prop. 1 \& 2]{PodlesSA} and also Appendix \ref{app}):
\begin{align}
\zeta_{\Delta_q\spin}(s) &= \Tr_{\H_q}\, (\Delta_q\spin)^{-s} = \Tr_{\H_q} \vert \DD_q \vert^{-2s} = \zeta_{\vert \DD_q \vert}(2s) \notag\\
&= 4u^{-2s} \sum_{k=1}^{\infty} k \frac{q^{2ks}}{(1 - q^{2k})^{2s}} = 4u^{-2s} q^{2s} \sum_{n=0}^{\infty} \frac{\Gamma(n + 2s)}{n!\, \Gamma(2s)} \frac{q^{2n}}{(1 - q^{2(n+s)})^2}\,.\label{z1}
\end{align} 

The last formula provides a valid meromorphic extension to the entire complex plane.

The full asymptotic expansion of $\Pspn(\sigma)$ could again be deduced from formula \eqref{z1} via the inverse Mellin transform, along the lines of \cite{PodlesSA}. The rather tedious computations can be bypassed by noting that $\Delta_q\spin$ and $\Delta_q\simp$ commute and differ by a bounded perturbation $\Delta_q\spin - \Delta_q\simp = - 2 + (\Delta_q\simp)^{-1}$. Consequently, we can write
\begin{align*}
\Psim(\sigma) - \Pspn(\sigma) & = \Tr \left( e^{-\sigma \Delta_q\simp} - e^{-\sigma \Delta_q\spin} \right) =  \Tr e^{-\sigma \Delta_q\simp} \left(1 - e^{-\sigma (\Delta_q\spin - \Delta_q\simp)} \right) \\
& \leq \big\Vert 1 - e^{-\sigma (\Delta_q\spin - \Delta_q\simp)} \big\Vert \, \Psim(\sigma) = \OO(\sigma \log^2 \sigma).
\end{align*}

The last equality follows from an operatorial inequality $\lim_{t\to 0} \tfrac{1}{t}\norm{1- e^{-tX}} \leq \norm{X}$ (see \cite[Remark 4.12]{PodlesSA}) and Formula \eqref{heat_q0}. This means that 
\begin{align}\label{Pspin}
\Pspn(\sigma) = \tfrac{1}{4 \log^2q}\, \big[ 2 \log^2 (u\sigma) + G\big(\log (u\sigma) \big)\, \log (u\sigma) + F\big(\log(u\sigma)\big) \big] + R_{\rm sp}(u\sigma)\,.
\end{align}
Hence indeed the leading small-$\sigma$ behaviour of $\Pspn(\sigma)$ is captured by Formula \eqref{heat_q0} for the simplified Laplacian. This harmonises with the fact that the $n=0$ term in the last formula in \eqref{z1} is nothing but $\zeta_{\Delta_q\simp}$.

The structure of the remainder $R_{\rm sp}(\sigma)$ can be inferred from Formula \eqref{z1} for the zeta function, in close analogy with \cite[Theorem 4.4]{PodlesSA}. Observe that $\zeta_{\Delta_q\spin}$ has poles located on a regular lattice in the left complex half-plane (cf. Figure \ref{fig:QS_Sd} c)). Consequently, $\Gamma \cdot \zeta_{\Delta_q\spin}$ has third order poles at negative integers --- yielding $\sigma^n \log^2\sigma$ contribution and double poles elsewhere --- giving rise to $\sigma^{\pi i j / \log q} \sigma^n \log\sigma$ oscillatory terms. Summa summarum,
\begin{align}\label{Rspin}
R_{\rm sp}(x) \, \tzero \, \sum_{n = 1}^{\infty} \left[ h_n \log^2 x + G_n(\log x) \log x + F_n(\log x) \right] x^{n},
\end{align}
where $h_n \in \sR$ and $F_n, G_n$ are periodic bounded functions of the form similar to that of $F$ and $G$. Let us stress that the sum over $n$ need not a priori to be convergent, which is just a restatement of the fact that asymptotic expansions of heat traces are generically divergent.

This analysis leads to the conclusion about the dimension spectrum:
\begin{align*}
\Sd(\Delta_q\spin) = \tfrac{\pi i}{\log q} \sZ - \sN = \{ \tfrac{\pi i}{\log q} k - n\, |\, k \in \sZ, n \in \sN\}\,, \quad \text{with} \quad \ord\Sd = 3\,.
\end{align*}
Observe the difference with respect to \eqref{SimSpec}, illustrated in Fig. \ref{fig:QS_Sd}.

Even though we only have an asymptotic formula for $\Pspn$, we can deduce the leading behaviour of the spectral dimension associated with the spinor Laplacian. This is because formula \eqref{Rspin} grants us an explicit control on the remainder. We thus have
\begin{align*}
d_S^{\,q,\text{sp}}(\sigma) & = -2 \, \frac{\big[ G'\big(\log (u \sigma)\big) + 4 \big] \log (u \sigma) + F'\big(\log (u \sigma) \big) + G\big(\log (u \sigma) \big)}{2 \log^2 (u \sigma) + G \big(\log (u \sigma) \big) \log (u \sigma) + F\big(\log (u \sigma) \big)} + \OO(\sigma) \\
& = d_S^{\,q,\text{sm}}(\sigma) + \OO(\sigma)\,.
\end{align*}
Hence, also the spectral dimensions associated with the spinor and simplified Laplacians on the quantum sphere share the same leading behaviour for small $\sigma$ --- see Figure \ref{fig:simp1}.

\subsection{Scalar Laplacian \label{subsec:scalar}}

The `scalar Laplacian' $\Delta_q\scl$, introduced in \cite{Benedetti:2009fe}, originates from the Casimir operator on the Hopf algebra $\mathcal{U}_q(\mathfrak{su}(2))$. It acts on a Hilbert space $\H_q'$  (on which the algebra $\A_q$ can also be faithfully represented) spanned by orthonormal vectors $\ket{j,m}$, with $m \in \{-j,-j+1,\ldots,j\}$ and $j \in \sN$ as
\begin{align}
\Delta_q\scl \, \ket{j,m} & = \frac{\cosh\big( \tfrac{1}{2} (2j+1) \log q\big) - \cosh \big(\tfrac{1}{2} \log q \big)}{2 \sinh^2 \big(\tfrac{1}{2} \log q \big)} \, \ket{j,m} \notag \\
& = u(\sqrt{q}) \, q^{-1/2} \left( q^{-j} - 1 - q + q^{j+1} \right) \ket{j,m}\,. \label{Delta_scal}
\end{align}

In the limit $q \to 1$ the operator $\Delta_q\scl$ tends (strongly) to the standard scalar Laplacian on $S^2$ (cf. \eqref{S2scal}). In contradistinction with $\Delta_q\spin$, it does have a zero mode, which means that $\Pscl(\sigma)$ tends to $\dim \Ker\Delta_q\scl = 1$ as $\sigma$ goes to infinity.

The small-$\sigma$ behaviour of the heat trace can be deduced by singling out the unbounded part of $\Delta_q\scl$, as in the case of $\Delta_q\spin$. Namely, we have
\begin{align*}
\Pscl (\sigma) & = \Tr_{\H_q'} e^{- \sigma \Delta_q\scl} = \sum_{j=0}^{\infty} (2j+1) \exp \left\{ - \sigma \, u(\sqrt{q}) \, q^{-1/2} \left( q^{-j} - 1 - q + q^{j+1} \right) \right\} \\
& = 1 + \sum_{n=1}^{\infty} (2n+1) \exp \left\{ - \sigma \, u(\sqrt{q}) \, q^{-1/2} \left( q^{-n} - 1 - q + q^{n+1} \right) \right\} \\
& = 1 + \sum_{n=1}^{\infty} (2n+1) \exp \left\{ - \sigma \, u(\sqrt{q}) \, q^{-1/2} q^{-n} \right\} + \OO(\sigma \log^2 \sigma) \\
& = 1 + \tfrac{1}{2} \P_{\sqrt{q}}\simp(\sigma q^{-1/2}) + \sum_{n=1}^{\infty} \exp \left\{ - \sigma \, u(\sqrt{q}) \, q^{-1/2} q^{-n} \right\} + \OO(\sigma \log^2 \sigma).
\end{align*}

The last series can be evaluated explicitly \cite[Proposition 12]{HeatEZ} using the inverse Mellin transform technique:
\begin{align*}
\sum_{n=1}^{\infty} e^{-x q^{-n}} = \tfrac{1}{\log q} \left[ \log x + \gamma - \tfrac{1}{2} \log q + H(\log x) \right] + \OO(x)\,,
\end{align*}
with
\begin{align*}
H(x) &\vc - \sum_{k \in \sZ^*} \Gamma(-\tfrac{2 \pi i}{\log q}\, k)\, e^{2 \pi i k x/ \log q}\,.
\end{align*}

This yields
\begin{align}\label{Pscl}
\Pscl (\sigma) & = \tfrac{1}{2} \, \P_{\sqrt{q}}\simp(\sigma q^{-1/2}) + \tfrac{1}{\log q} \big[ \log (\sigma \, u(\sqrt{q})) + \gamma + H\big( \log (\sigma \, u(\sqrt{q}) \, q^{-1/2}) \big) \big] + \OO(\sigma \log^2 \sigma)\,.
\end{align}

Similarly as in the spinor case, one can unfold the structure of the remainder by an inspection of the zeta function $\zeta_{\Delta_q\scl}$ --- see Appendix \ref{app}. Its meromorphic structure is very similar to that of $\zeta_{\Delta_q\spin}$. The conclusion is that $\Delta_q\scl$ and $\Delta_q\spin$ have identical dimension spectra, both of order 3.

Formula \eqref{Pscl} allows us also to compute the spectral dimension associated with $\zeta_{\Delta_q\scl}$ up to the terms of order $\OO(\sigma)$ at $\sigma = 0$
\begin{align}
d_S^{\,q,\text{sc}}(\sigma) & = \frac{\tfrac{1}{2} \sigma q^{-1/2} \, (\P_{\sqrt{q}}\simp)'(\sigma q^{-1/2}) + \tfrac{1}{\log q} \big[ 1 + H'\big( \log (\sigma \, u(\sqrt{q}) \, q^{-1/2}) \big) \big]}{\tfrac{1}{2} \, \P_{\sqrt{q}}\simp(\sigma q^{-1/2}) + \tfrac{1}{\log q} \big[ \log (\sigma \, u(\sqrt{q})) + \gamma + H\big( \log (\sigma \, u(\sqrt{q}) \, q^{-1/2}) \big) \big]} + \OO(\sigma) \notag \\
& = d_S^{\sqrt{q},\text{sm}}(q^{-1/2} \sigma) + \OO\big( (\log\sigma)^{-2} \big)\,, \label{sc_sm}
\end{align}
where the last equality follows from the exact formula \eqref{dSsimp}.
It shows that the leading small-$\sigma$ behaviour of the spectral dimension for the scalar Laplacian is captured by the rescaled spectral dimension for $\Delta_q\simp$.

\subsection{Comparison \label{subsec:QS_comp}}

We now compare the results obtained for the three Laplacians and contrast them with their classical counterparts.

Recall that the spinor Laplacian\footnote{In the mathematical literature \cite{Friedrich} the ``spinor Laplacian'' $\Delta^\mathcal{S}$ is slightly different from $\Delta\spin$, which is the square of the Dirac operator $\Dslash^2$. The two operators are related through the Schr\"odinger--Lichnerowicz formula $\Dslash^2 = \Delta^\mathcal{S} + \tfrac{1}{4} \R$, with $\R$ being the scalar curvature. On the unit 2-sphere with a round metric the difference amounts to a trivial shift $\Delta\spin = \Dslash^2 = \Delta^\mathcal{S} + 2$.} acts on spinor harmonics over the unit two-sphere as follows \cite{Trautman,BarHomogeneous}:
\begin{align}\label{S2spin}
\Delta\spin \ket{l,m}_\pm = (l + \tfrac{1}{2})^2 \ket{l,m}_\pm\,.
\end{align}
In turn, for the scalar Laplacian (a.k.a. the Laplace--Beltrami operator) we have \cite{Shubin}:
\begin{align}\label{S2scal}
\Delta\scl \ket{j,m} = j (j+1) \ket{j,m}\,.
\end{align}

\begin{figure}[h]
\centering
\includegraphics[width=0.3\textwidth]{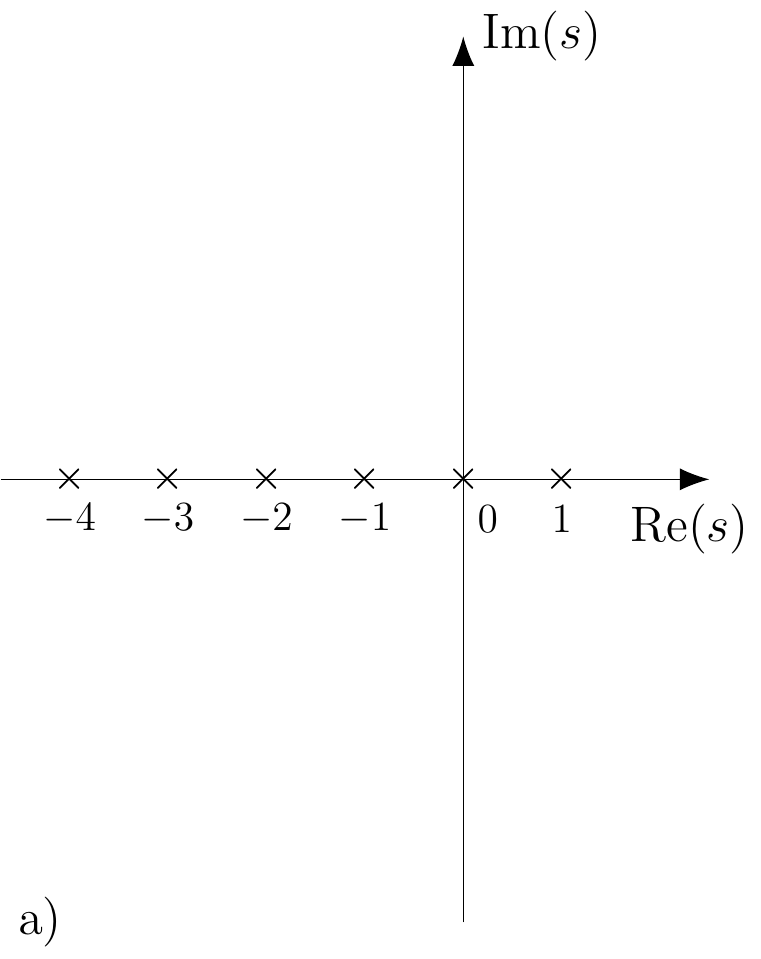}
\hspace{0.02\textwidth}
\includegraphics[width=0.3\textwidth]{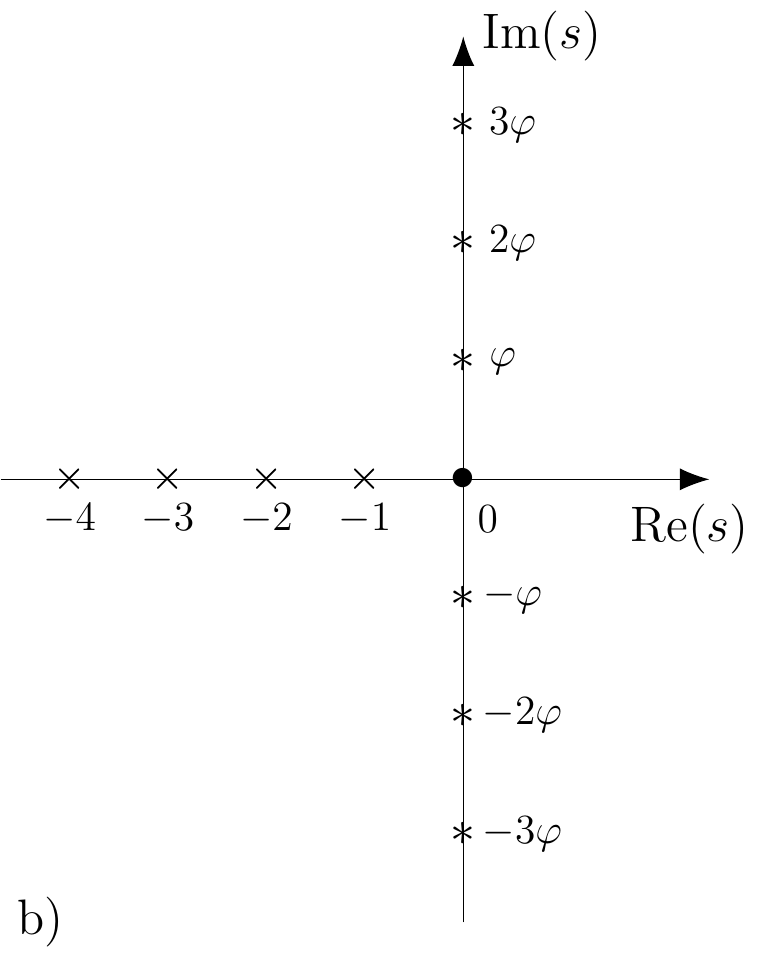}
\hspace{0.02\textwidth}
\includegraphics[width=0.3\textwidth]{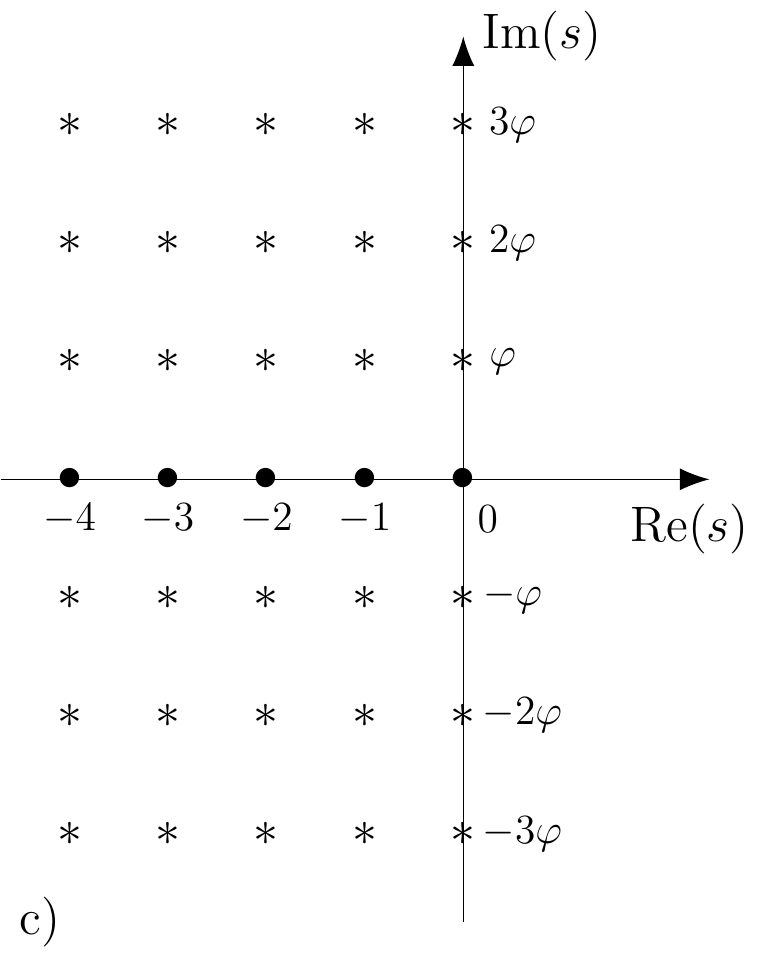}
\caption{\label{fig:QS_Sd}A comparison of dimension spectra for different Laplacians on 2-sphere $\Sd(\Delta\spin) = \Sd(\Delta\scl)$  (plot a)) and on quantum sphere  $\Sd(\Delta_q\simp)$ (plot b)), $\Sd(\Delta_q\spin) = \Sd(\Delta_q\scl)$ (plot c)). The symbols $\times$, $\ast$ and $\bullet$ denote points in $\Sd$, corresponding to poles of the function $\Gamma \cdot \zeta$ of order 1, 2 and 3 respectively, while $\varphi = \pi / (\log q)$. An $n$-th order pole of the function $\Gamma \cdot \zeta$ yields a term proportional to $(\log \sigma)^{n-1}$ in the asymptotic expansion of $\P(\sigma)$ (recall Eq. \eqref{poles}).}
\end{figure}

Let us firstly have a look at the dimension spectra at Fig. \ref{fig:QS_Sd}. Both $\Delta\spin$ and $\Delta\scl$ are classical differential operators of second order acting over a two-dimensional manifold. Consequently, we have \cite{Gilkey2} (see also \cite{ET_NEW} for a direct computation):
\begin{align}
\Sd(\Delta\spin) = \Sd(\Delta\scl) = 1 - \N\,, \quad \text{with} \quad \ord \Sd = 1\,. 
\end{align}

In the classical case we have $d_{\Sd} = 1$, in agreement with the general formula $d_{\Sd} = d /\eta$, as $d=2$ and $\eta =2$. In contrast, for all three operators on the quantum sphere we have $d_{\Sd} = 0$. In order to interpret this fact as the dimension drop, i.e. $d=0$, we need to argue that the `quantum Laplacians' are of order $\eta > 0$.

In the framework of spectral triples, the operator $\D_q$ verifies the so-called first-order condition \cite{dab_sit}, which mimics the demand for a classical Dirac operator to be a first order differential operator \cite{ConnesNCG,Varilly}. On the physical side, this condition limits the admissible fluctuations of an operator $\D$ by gauge fields \cite{ConnesFirst} and thus it is pertinent in building particle physics models from noncommutative geometry \cite{WalterBook,ConnesFirstSM}. Since $\D_q$ is a first-order operator, for $\Delta_q\spin = \D_q^2$ we should set $\eta = 2$. The operator $\D_q^S$ does not meet the first-order condition \cite{PodlesSA} and $\Delta_q\scl$ does not come from a `Dirac' operator at all. Nevertheless, $\Delta_q\simp$  differs from $\Delta_q\spin$ by a bounded perturbation and so does $\Delta_q\scl$ after a suitable rescaling and reparametrisation $q \rightsquigarrow \sqrt{q}$. We can thus safely assume that $\eta(\Delta_q\scl) = \eta(\Delta_q\simp) = 2$ and conclude that indeed the dimension of the quantum sphere is 0. The issue of the order of an operator over a quantum space is more subtle for the $\kappa$-Minkowski space, as we will see in the next section.

The existence of non-positive numbers in the dimension spectra of the two classical Laplacians on the two-sphere certify the impact of the non-trivial Riemann tensor on the heat trace (recall Formula \eqref{heat_pdo}). The dimension spectra of quantum Laplacians also contain negative numbers, which suggest that quantum spheres are `curved' in some sense (see e.g. \cite{ConnesMarcolli,ConnesModular,Paolo2005,FatKhal,JMP2016,SitarzNonminimal} for a discussion of curvature of quantum spaces). On top of that, these dimension spectra contain points outside of the real axis, which hint at some kind of self-similar structure of the quantum sphere \cite{Kellendonk,PodlesSA} (see also \cite{dunne2012,DunnePRL}). Note also that, excluding the simplified case of $\Delta_q\simp$, the non-real points appear all over the left complex-half plane. This suggests that the curvature and self-similar structure of a quantum sphere are deeply interwoven.

Finally, the dimension spectra of quantum Laplacians are of order three, which means that they are already beyond the realm of classical pseudodifferential operators, as the latter can only have $\ord \Sd = 2$. Third order poles in the dimension spectra have been detected in the context of manifolds with conical singularities \cite{Lescure}. They occur when one studies the conical singularities (and, more generally, stratified spaces) from the perspective of manifolds with boundary \cite{LeschFuchs}. Concretely, specific non-local boundary conditions related to the singularity coerce the use of Fuchs-type operators \cite{Gil2008}. Although the mathematical context here is very different, one might take it as a (not so surprising) indication that the geometry of the quantum sphere is not smooth.

\medskip

We now turn to the analysis of the spectral dimensions.

\begin{figure}[h]
\includegraphics[width=0.45\textwidth]{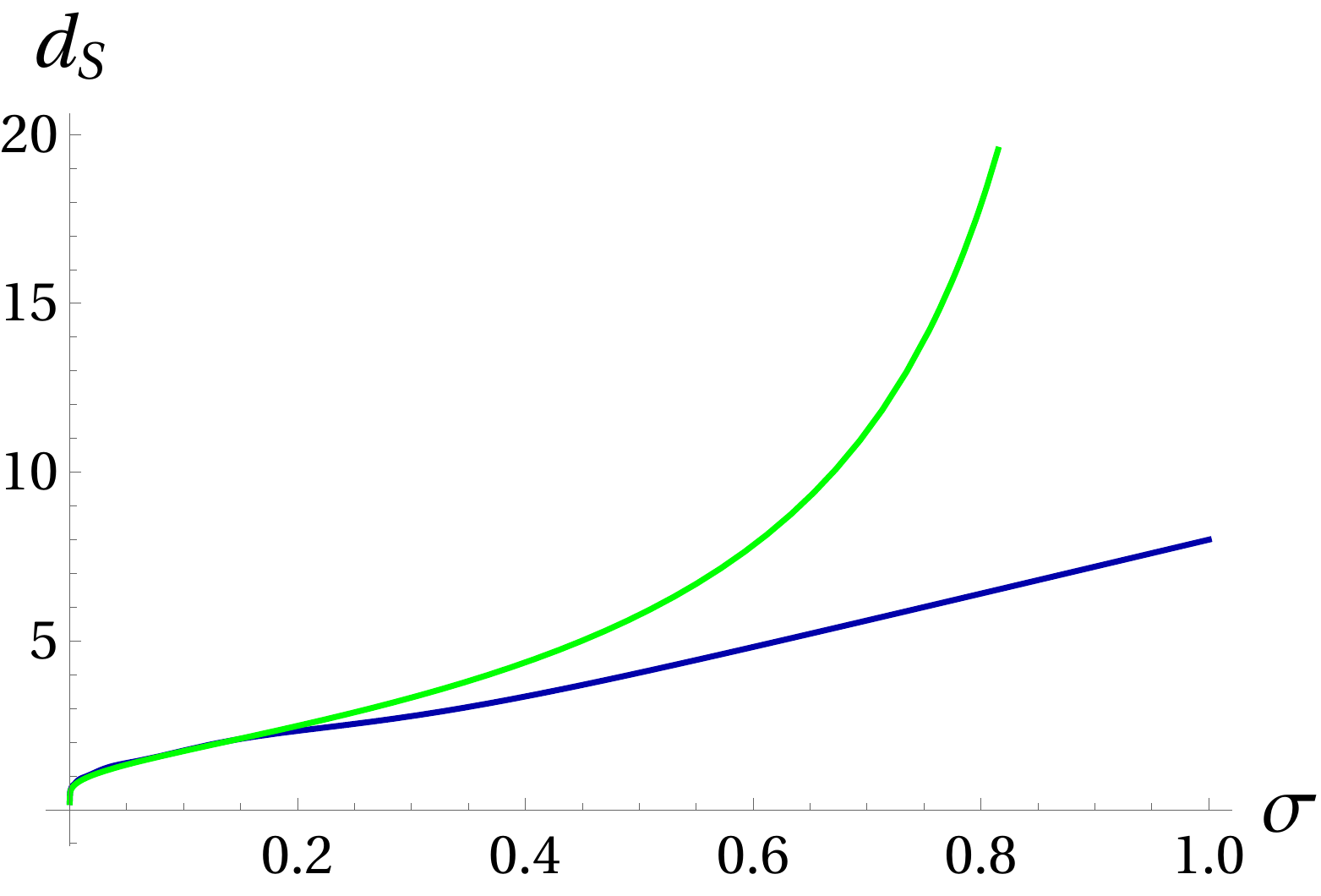}
\hspace{0.05\textwidth}
\includegraphics[width=0.45\textwidth]{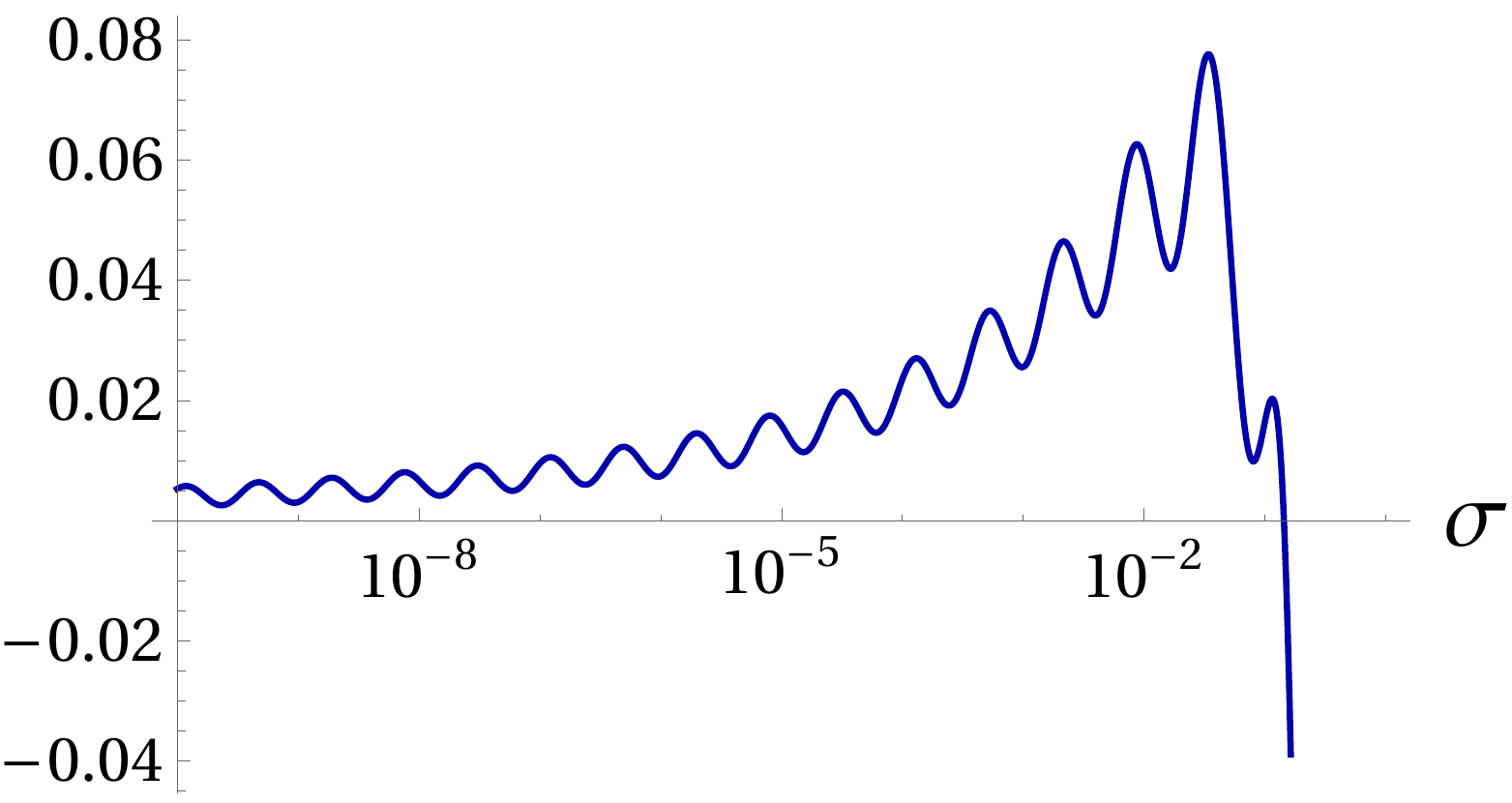}
\caption{\label{fig:simp1} (LHS) The dark blue line is the spectral dimension for the Laplacian $\Delta_q\simp$, computed numerically from Eq. \eqref{dS_eig} up to the 100th eigenvalue. The leading behaviour in the UV is determined from Formula \eqref{dS_simp_lead} as $-4/ (\log \sigma)$ (light green line). (RHS) The function $d_S^{\,q, \text{sm}}$ after subtraction of the leading behaviour $-4/ (\log \sigma)$ clearly shows the $\log$-periodic oscillations.}
\end{figure}

Let us first have a look at the small-$\sigma$ behaviour of $d_S(\sigma)$ for the simplified Laplacian $\Delta_q\simp$. From Eq. \eqref{dSsimp} one deduces the leading behaviour in the UV:
\begin{align}\label{dS_simp_lead}
d_S^{\,q, \text{sm}}(\sigma) = \frac{-4}{\log \sigma} \left ( 1 + \frac{\pi i}{\log q} \sum_{j \in \sZ^*} j \, \Gamma(\tfrac{\pi i}{\log q}\, j)\, \sigma^{-\pi i j/ \log q}\right) + \OO((\log \sigma)^{-2})\,.
\end{align}
The function $d_S^{\,q, \text{sm}}$ for $q = 1/2$ is illustrated on Fig. \eqref{fig:simp1}.

\begin{figure}[h]
\includegraphics[width=0.5\textwidth]{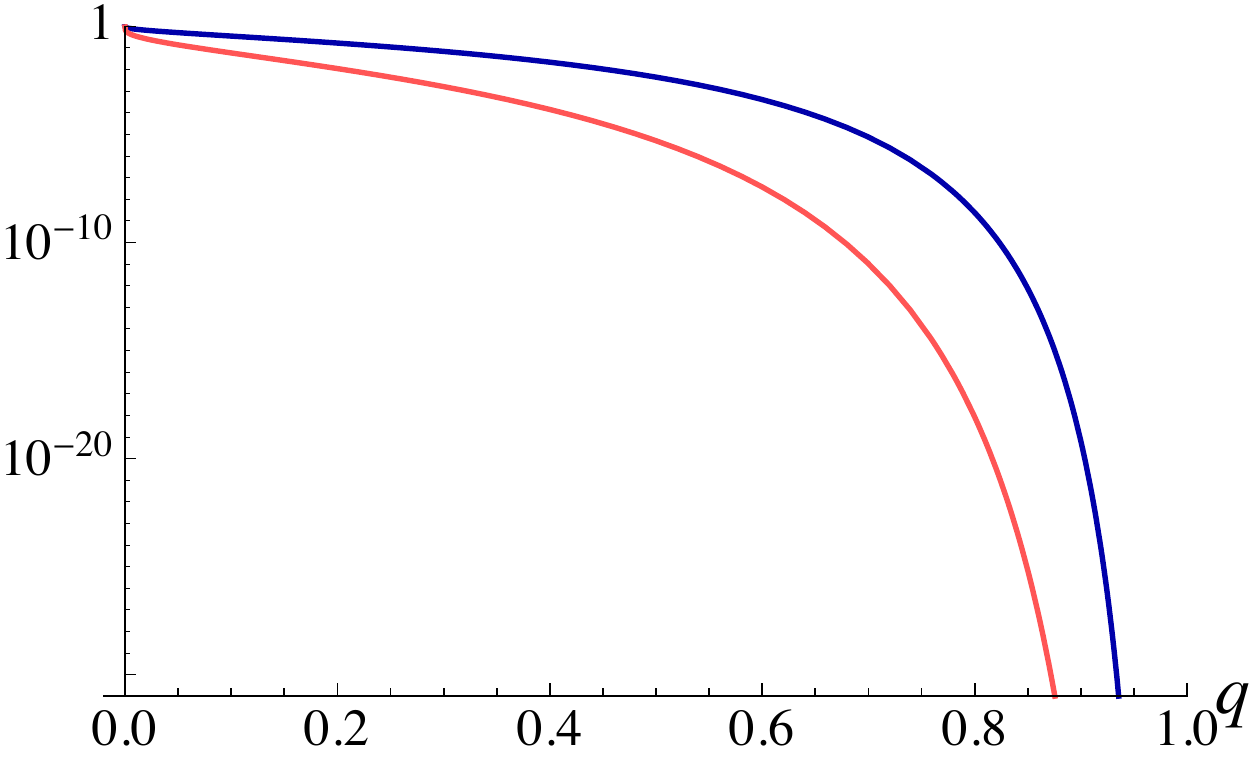}
\caption{\label{fig:Amp} The $q$-dependence of the amplitude of (leading frequency) oscillations in the spectral dimension for $\Delta_q\spin$ (dark blue) and $\Delta_q\scl$ (light red). The former is given by the function $A(q)$ defined in \eqref{amp}, whereas the latter is $A(\sqrt{q})$, because of the relation \eqref{sc_sm}.}
\end{figure}

Formula \eqref{dS_simp_lead} shows that the spectral dimension of $\Delta_q\simp$ drops to zero in the UV. Observe that the slope $-4/ (\log \sigma)$ does not depend on $q$. On the other hand, the amplitude of oscillations exhibits strong dependence on the value of $q$. Concretely, for the leading frequency ($j=1$ in the sum in Eq. \eqref{dS_simp_lead}) we have
\begin{align}\label{amp}
A(q) = \big\vert \tfrac{\pi}{\log q} \, \Gamma(\tfrac{\pi i}{\log q}) \big\vert.
\end{align}
The amplitude of oscillations tends to 1 as $q \to 0$ and decays very rapidly, as $e^{\pi^2 /( 2(q-1))}$, when $q$ goes to 1 --- see Fig. \ref{fig:Amp}.

\begin{figure}[h]
\includegraphics[width=0.45\textwidth]{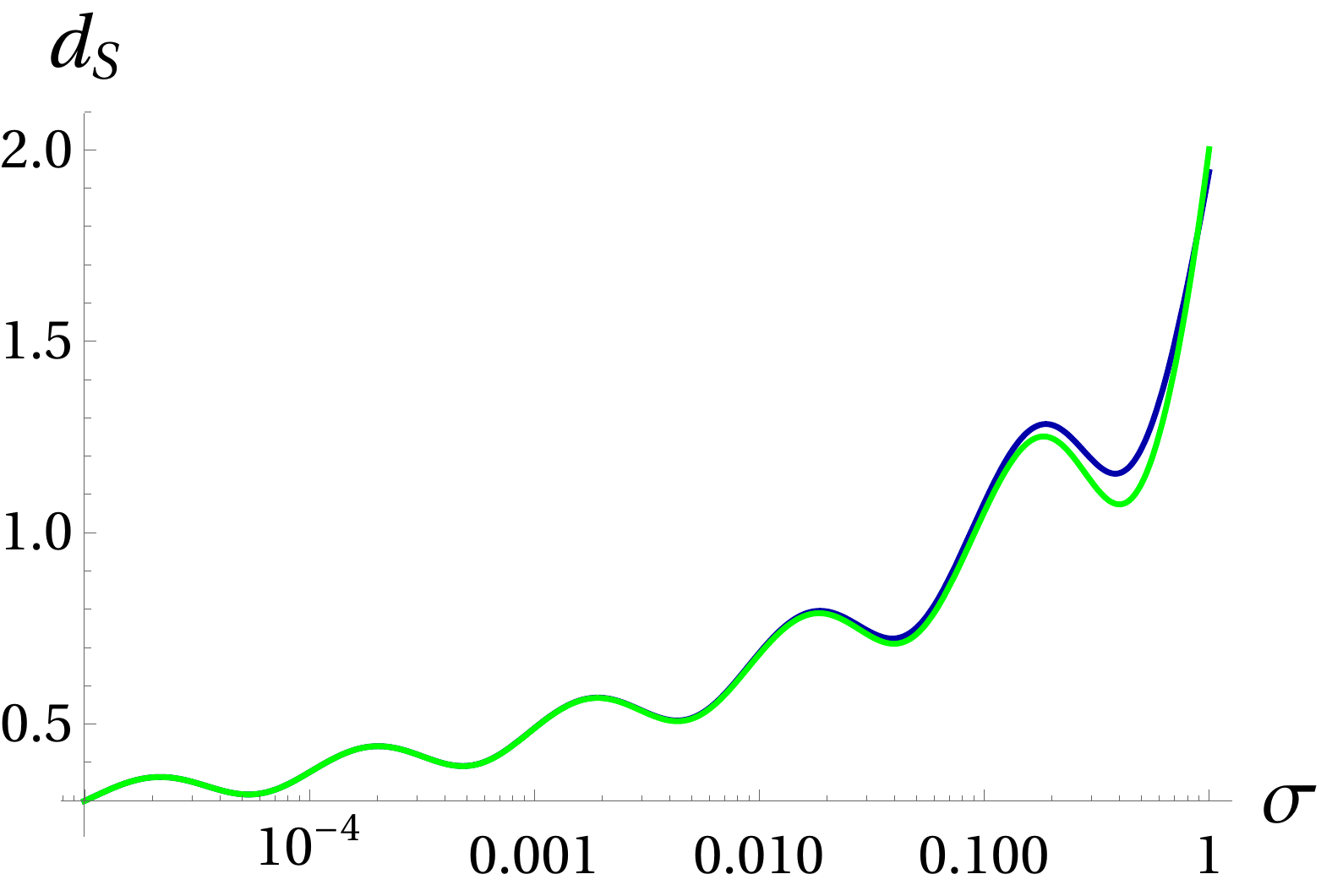}
\hspace{0.05\textwidth}
\includegraphics[width=0.45\textwidth]{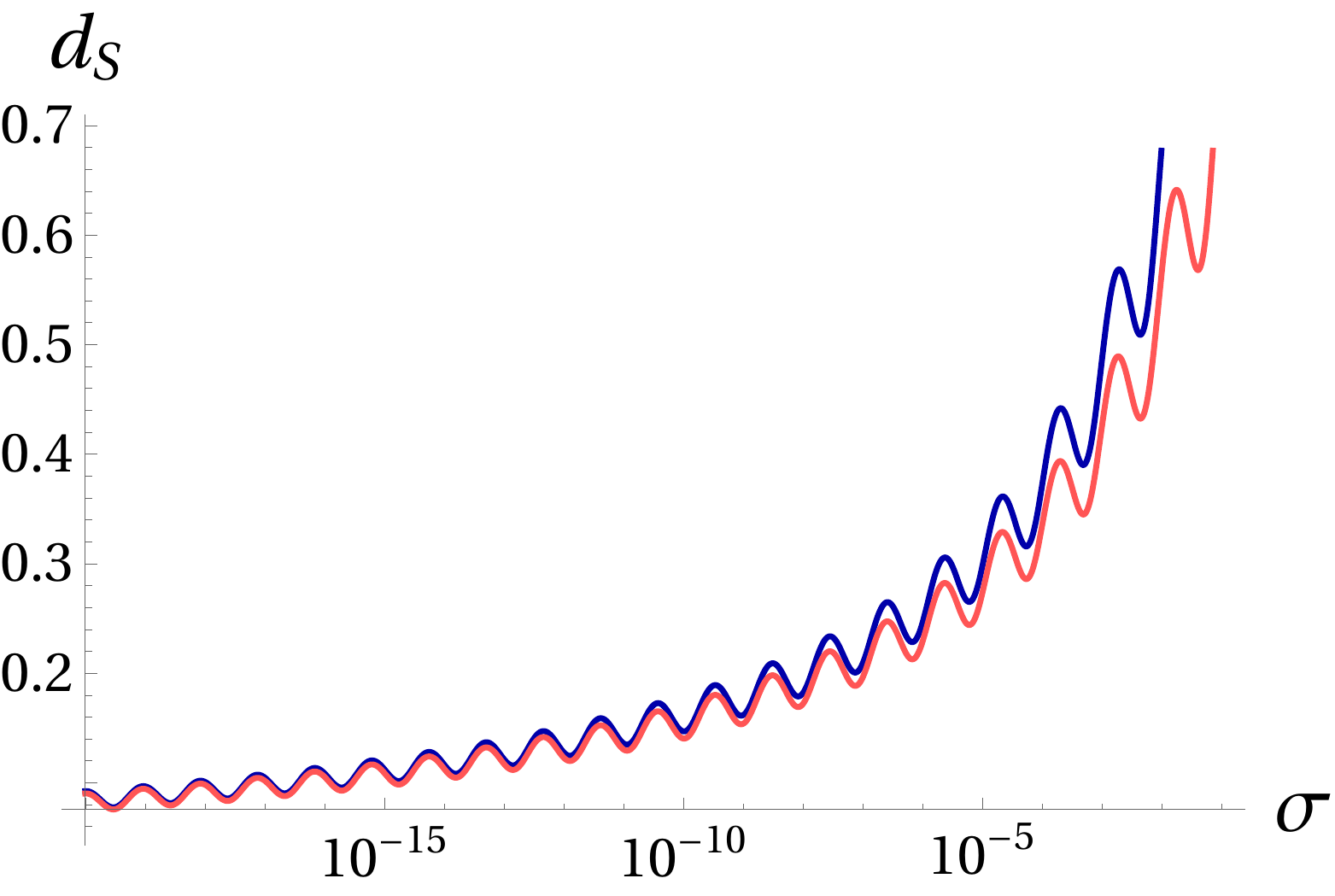}
\caption{\label{fig:QS_comp} A comparison of the UV behaviour of the spectral dimensions for three Laplacians on the the quantum sphere for $q=1/3$. On LHS the dark blue line corresponds to $d_S^{\,q, \text{sm}}(\sigma)$ and the light green one to $d_S^{\,q, \text{sp}}(\sigma)$. On the RHS the dark blue line is $d_S^{\,q, \text{sm}}(\sigma)$, whereas the light red one shows $d_S^{\,q^2, \text{sc}}(q \sigma)$.}
\end{figure}

In the previous sections we have shown that the leading UV behaviour of the spectral dimension for the spinor Laplacian is captured by $d_S^{\,q, \text{sm}}$. The same is true, after suitable rescaling, also for the scalar Laplacian, though with worse precision. The situation is illustrated on Fig. \ref{fig:QS_comp} through the numerical summation in Eq. \eqref{dS_eig} up to 100th eigenvalue.

\begin{figure}[h]
\includegraphics[width=0.5\textwidth]{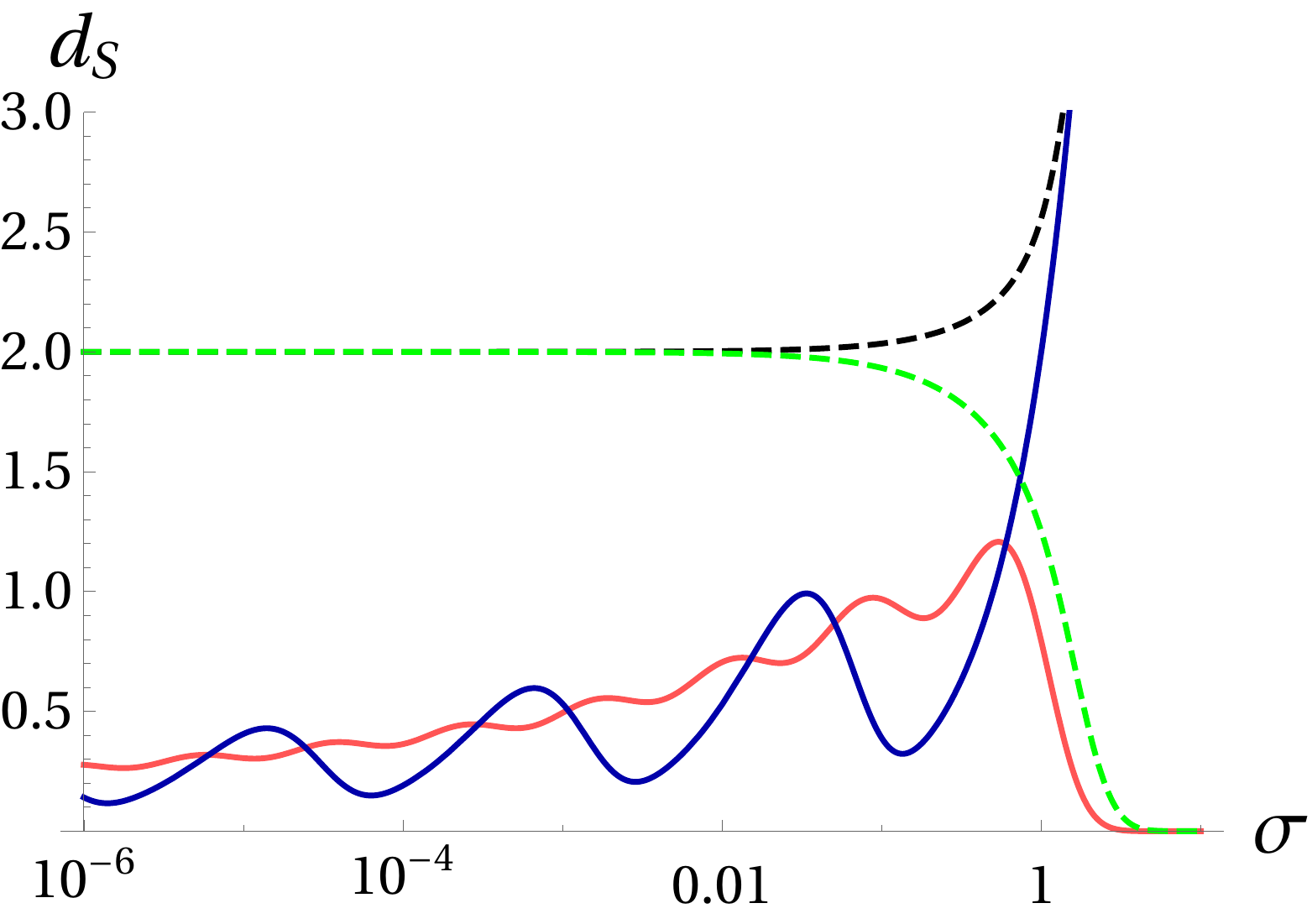}
\caption{The spectral dimension $d_S(\sigma)$ for $\Delta_q^{\rm sc}$ (continuous decaying curve) and $\Delta_q^{\rm sp}$ (continuous diverging) with $q = 0.15$, and for $S^2$ with the scalar (dashed decaying curve) and spinorial (dashed diverging) Laplacians.
\label{fig:ddqs}}
\end{figure}

\begin{figure}[h]
\includegraphics[width=0.5\textwidth]{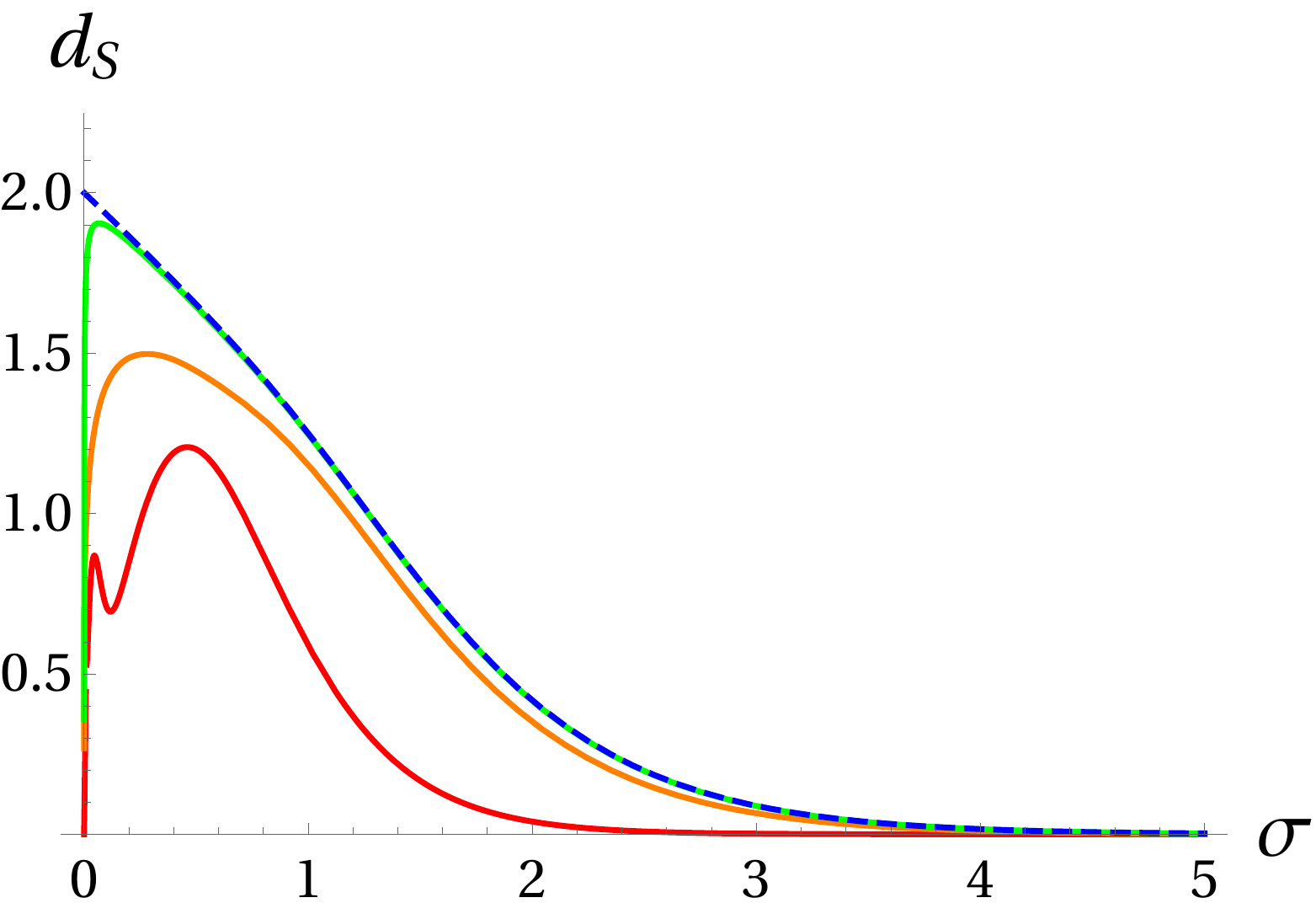}
\caption{The spectral dimension $d_S(\sigma)$ for $\Delta_q^{\rm sc}$, with $q = 0.1$ (bottom continuous curve), $q = 0.5$ (middle continuous) and $q = 0.9$ (top continuous), and for $S^2$ with the scalar Laplacian (dashed)
\label{fig:ddqs2}}
\end{figure}

A comparison of the spectral dimension associated with the operators $\Delta_q\spin, \Delta_q\scl$ and their classical counterparts $\Delta\spin, \Delta\scl$, respectively, is presented on Figs. \ref{fig:ddqs} and \ref{fig:ddqs2}. The main conclusion is that the spectral dimensions of both quantum Laplacians on the quantum sphere exhibit log-periodic oscillations in UV. The amplitude of these oscillations drops very rapidly when $q$ tends to the classical value 1 (cf. Fig. \ref{fig:Amp}). In particular, for $\Delta_q\scl$ the value $q = e^{-0.01} \approx 0.99$ adopted on Fig. 1 in \cite{Benedetti:2009fe}, we have $A(\sqrt{q}) \sim 10^{-430}$, which explains why the oscillations have been overlooked in that paper.


\newpage
\section{\texorpdfstring{$\kappa$-Minkowski spacetime}{k-Minkowski spacetime}} \label{sec:kMink}

The $\kappa$-Minkowski space was introduced \cite{Majid:1994by} in the context of quantum groups that describe deformed relativistic symmetries -- it is the spacetime that is bi-covariant under the action and coaction (the former is determined by the algebraic and the latter by the coalgebraic structure) of the $\kappa$-Poincar\'{e} Hopf algebra. Both these mathematical objects can be defined in any number of dimensions \cite{Lukierski:1994qn} but $\kappa$-Poincar\'{e} algebra was first derived \cite{Lukierski:1991qa,Lukierski:1992ny} in the (3+1)d case, as a contraction of the quantum-deformed anti-de Sitter algebra $U_q(\mathfrak{so}(3,2))$, obtained by taking the limit of the (real) deformation parameter $q \rightarrow 1$ and anti-de Sitter radius $R \rightarrow \infty$, while their ratio $R\, \log q = \kappa^{-1}$, $\kappa > 0$ is kept fixed. Hence the new deformation parameter $\kappa$ has the dimension of inverse length (in contrast to the dimensionless $q$), which allows for the geometrization of Planck mass $m_P = \hbar c^{-1} \lambda_P$, expressed in terms of Planck length $\lambda_P$. This peculiar feature enabled application of $\kappa$-Poincar\'{e} algebra in the construction of models of so-called doubly special relativity, which was subsequently recast as the relative locality framework and serves as an important source for the quantum gravity phenomenology \cite{Amelino:2013qy}. The interest in $\kappa$-Poincar\'{e} and $\kappa$-(anti-)de Sitter algebras is also motivated by results for the 2+1-dimensional gravity, where it is known that they may arise from the structure of classical theory, at least in certain special cases (see \cite{Kowalski:2019qs} and references therein).

The $n\!+\!1$-dimensional $\kappa$-Minkowski noncommutative space is the dual of the subalgebra of translations of the ($n\!+\!1$-dimensional) $\kappa$-Poincar\'{e} algebra, since the latter is naturally interpreted as the algebra of spacetime coordinates. The time $X_0$ and spatial coordinates $X_a$, $a = 1,...,n$ satisfy the following commutation relations
\begin{align}\label{kMink_alg}
[X_0,X_a] = \frac{i}{\kappa}\, X_a\,, \qquad [X_a,X_b] = 0\,.
\end{align}
As a vector space, the $\kappa$-Minkowski space is isomorphic to the ordinary Minkowski space in $n\!+\!1$ dimensions, which can be recovered in the classical limit $\kappa \rightarrow \infty$. The Lie algebra generated by $X_0,X_a$ is usually denoted $\mathfrak{an}(n)$, where the notation refers to $n$ Abelian and nilpotent generators $X_a$. 

The corresponding Lie group ${\rm AN}(n)$ has the geometry of $n\!+\!1$-dimensional elliptic de Sitter space \cite{Kowalski:2002dy,Kowalski:2003de}. Group elements can be written as ordered exponentials, whose ordering is equivalent to the choice of coordinates on the group manifold. For example, in the time-to-the-right ordering a group element has the form
\begin{align}
g = e^{-i P^a X_a} e^{i P_0 X_0}
\end{align}
and $P_0,P_a \in \sR$ are coordinates in the so-called bicrossproduct basis. It is easy to notice that we can interpret ${\rm AN}(n)$ as momentum space if such exponentials are treated as plane waves on $n\!+\!1$-dimensional $\kappa$-Minkowski space. ${\rm AN}(n)$ is equipped with the structure of the algebra of translations and is related with spacetime coordinate algebra via the group Fourier transform \cite{Freidel:2007fe,Freidel:2008fy}. Furthermore, the geometry of momentum space becomes evident in the classical basis (the name refers to the classical form of expression for the dispersion relation in this case), which can be introduced via the transformation
\begin{align}\label{eq:4.5a}
p_0 & = \kappa \sinh\left(\tfrac{P_0}{\kappa}\right) + \tfrac{1}{2\kappa} e^{P_0/\kappa} P_aP^a\,, \nonumber\\ 
p_a & = e^{P_0/\kappa} P_a\,, \nonumber\\ 
p_{-1} & = \kappa \cosh\left(\tfrac{P_0}{\kappa}\right) - \tfrac{1}{2\kappa} e^{P_0/\kappa} P_aP^a\,.
\end{align}
The above relations lead to the constraints on $\{p_0,p_a,p_{-1}\}$, $-p_0^2 + p_ap^a + p_{-1}^2 = \kappa^2$ and $p_0 + p_{-1} > 0$, which describe the embedding of ${\rm AN}(n)$ as half of a $(n,1)$-hyperboloid in $(n\!+\!1)\!+\!1$-dimensional Minkowski space. $p_{-1}$ is just an auxiliary coordinate, diverging in the $\kappa \rightarrow +\infty$ limit. 

In order to consider the diffusion process (i.e. study the heat trace) determined by a Laplacian on $\kappa$-Minkowski space, one first has to perform the Wick rotation $(p_0 \mapsto i p_0,p_{-1} \mapsto i p_{-1},\kappa \mapsto i \kappa,P_0 \mapsto i P_0)$. This leads to the Euclidean momentum space, which has the hyperbolic geometry \cite{Arzano:2014de}. 

The Euclidean version of $\kappa$-Minkowski space has also been studied from the perspective of spectral triples \cite{IMS2012,MatassaMink,Iochum-Masson} via the star-product realisations \cite{meljanac2008,DurhuusSitarz,Pachol2015}. In this framework the algebra \eqref{kMink_alg} is faithfully represented on the Hilbert space $L^2(\sR^{n+1})$ through a left regular group representation.

As mentioned in the introduction, there are several possible choices for a Laplacian on (Euclidean) $\kappa$-Minkowski space. All of them have continuous spectra, what reflects the non-compactness of the underlying space. Therefore, in order to define the corresponding heat traces, one firstly needs to choose an IR cut-off. A natural choice is a function $f$ compactly supported on $\sR^{n+1}$ and promoted to an operator on $L^2(\sR^{n+1})$ through the Weyl-like quantisation (cf. \cite{DurhuusSitarz,MatassaMink}). Fortunately, it turns out that the regularising function factors out in trace formulae and contributes just a multiplicative factor $\norm{f}_{L^2}$ \cite{MatassaMink,Iochum-Masson}. This means that the return probability does not depend on the choice of the IR regularisation and can be computed via the heat kernel formula \eqref{eq:3.6}. In the classical basis it reads \cite{Arzano:2014de} \footnote{The factor $\kappa$ in the numerator was missing in \cite{Arzano:2014de} but it did not affect results for the spectral dimension; here we have also rescaled the integrand by 2.}:
\begin{align}\label{eq:4.5}
{\cal P}(\sigma) = \int\! d^{n+1}p\ \frac{\kappa}{\sqrt{p_0^2 + p_ap^a + \kappa^2}}\, e^{-\sigma {\cal L}(p_0, \{p_a\})},
\end{align}
where ${\cal L}$ is the (Euclidean) momentum space-representation of a Laplacian.

In the following Subsections \ref{sec:4.1}-\ref{sec:4.3} we will calculate the spectral dimensions of $\kappa$-Minkowski space equipped with three distinct Laplacians, in 2, 3 and 4 topological dimensions for each of them. In this way we improve and extend the results obtained by one of us in \cite{Arzano:2014de}. Furthermore, we compute the complete dimension spectra of the relevant Laplacians.

\subsection{Bicovariant Laplacian} \label{sec:4.1}

Let us first consider the Laplacian determined by the bicovariant 
differential calculus on the $\kappa$-Minkowski space \cite{Sitarz:1995ne,Gonera:1996de}, which we call the bicovariant Laplacian. In terms of bicrossproduct coordinates, the Euclidean momentum space representation of this Laplacian is given by
\begin{align}\label{eq:4.6}
{\cal L}_{\text{cv}}(P_0,\{P_a\}) = & \, 4\kappa^2 \sinh^2\left(\tfrac{1}{2\kappa} P_0\right) + e^{P_0/\kappa} P_a P^a \nonumber\\
&+ \frac{1}{4\kappa^2} \left(4\kappa^2 \sinh^2\left(\tfrac{1}{2\kappa} P_0\right) + e^{P_0/\kappa} P_a P^a\right)^2\,,
\end{align}
while in classical coordinates it acquires the familiar standard form
\begin{align}\label{eq:4.6a}
{\cal L}_{\text{cv}}(p_0, \{p_a\}) = p_0^2 + p_ap^a. 
\end{align}
The return probability in 3+1 dimensions reads \cite{Arzano:2014de}
\begin{align}\label{eq:4.7}
{\cal P}_{(3+1)}(\sigma) = \frac{\pi^2}{2\sigma^{3/2}} \left(2\kappa^2 \sqrt{\sigma} - \sqrt{\pi}\, e^{\kappa^2 \sigma} (2\kappa^2 \sigma - 1) \left(1 - {\rm erf}(\kappa \sqrt{\sigma})\right)\right)\,,
\end{align}
where ${\rm erf}(\cdot)$ is the error function, while in 2+1 dimensions we have
\begin{align}\label{eq:4.7a}
{\cal P}_{(2+1)}(\sigma) = \frac{\pi^{3/2} \kappa}{\sigma}\, U(\tfrac{1}{2},0;\kappa^2 \sigma)\,,
\end{align}
where $U(a,b;\cdot)$ is a Tricomi confluent hypergeometric function. Furthermore, we may also consider the case of 1+1 dimensions (not discussed in \cite{Arzano:2014de}),
\begin{align}\label{eq:4.7b}
{\cal P}_{(1+1)}(\sigma) = \frac{\pi^{3/2} \kappa}{\sqrt{\sigma}}\, e^{\kappa^2 \sigma} \left(1 - {\rm erf}(\kappa \sqrt{\sigma})\right)\,.
\end{align}

The exact formulae (\ref{eq:4.7}-\ref{eq:4.7b}) for heat traces can be directly developed into series around $\sigma = 0$
\begin{align}\label{eq:4.11a}
{\cal P}_{(3+1)}(\sigma) &= \frac{\kappa \pi^{5/2}}{2\sqrt{\sigma}^3} - \frac{\kappa^3 \pi^{5/2}}{2\sqrt{\sigma}} + \frac{4\kappa^4 \pi^2}{3} - \frac{3\kappa^5 \pi^{5/2}}{4} \sqrt{\sigma} + {\cal O}(\sigma)\,, \nonumber\\
{\cal P}_{(2+1)}(\sigma) &= \frac{2\kappa \pi}{\sigma} + \kappa^3 \pi \log\sigma + \kappa^3 \pi \left(1 + \gamma + 2 \log\frac{\kappa}{2}\right) + {\cal O}(\sigma)\,, \nonumber\\
{\cal P}_{(1+1)}(\sigma) &= \frac{\kappa \pi^{3/2}}{\sqrt{\sigma}} - 2\kappa^2 \pi + \kappa^3 \pi^{3/2} \sqrt{\sigma} + {\cal O}(\sigma)\,,
\end{align}
where $\gamma$ is the Euler--Mascheroni constant. From these expansions one can immediately read out the corresponding dimension spectra 
\begin{align}\label{eq:4.11b}
{\rm Sd}_{(3+1)} &= \{\tfrac{3}{2}\} \cup \{\tfrac{1-n}{2} \, \vert \, n \in \mathbb{N}\} = \{\tfrac{3}{2},\tfrac{1}{2},0,-\tfrac{1}{2},-1,-\tfrac{3}{2},\ldots\}\,, & \ord \Sd = 1\,, \nonumber\\
{\rm Sd}_{(2+1)} &= 1- \mathbb{N} = \{1,0,-1,-2,\ldots\}\,,  & \ord \Sd = 2\,, \nonumber\\
{\rm Sd}_{(1+1)} &= \tfrac{1}{2}(1- \mathbb{N}) = \{\tfrac{1}{2},0,-\tfrac{1}{2},-1,\ldots\}\,,  & \ord \Sd = 1\,.
\end{align}
Let us remind that in the 2+1 dimensional case the dimension spectrum is of the second order because of the presence of logarithmic terms in the expansion of ${\cal P}_{(2+1)}(\sigma)$.

We will now turn to results for the spectral dimensions.

From (\ref{eq:4.7}) one finds that in 3+1 dimensions the spectral dimension is given by the expression
\begin{align}\label{eq:4.8}
d_S^{(3+1)}(\sigma) = 3 + 2\kappa^2 \sigma \frac{2\kappa \sqrt{\sigma} - \sqrt{\pi}\, e^{\kappa^2 \sigma} (2\kappa^2 \sigma + 1) (1 - {\rm erf}(\kappa \sqrt{\sigma}))}{-2\kappa \sqrt{\sigma} + \sqrt{\pi}\, e^{\kappa^2 \sigma} (2\kappa^2 \sigma - 1) (1 - {\rm erf}(\kappa \sqrt{\sigma}))}\,,
\end{align}
which has the IR and UV limits $\lim_{\kappa\sqrt{\sigma} \rightarrow \infty} d_S^{(3+1)}(\sigma) = 4$ and $\lim_{\kappa\sqrt{\sigma} \rightarrow 0} d_S^{(3+1)}(\sigma) = 3$, respectively. Analogously, in 2+1 dimensions one uses (\ref{eq:4.7a}) to get
\begin{align}\label{eq:4.10}
d_S^{(2+1)}(\sigma) = 2 + \frac{\kappa^2 \sigma\, U(\frac{3}{2},1,\kappa^2 \sigma)}
{U(\frac{1}{2},0,\kappa^2 \sigma)}\,,
\end{align}
whose IR and UV limits are $\lim_{\kappa\sqrt{\sigma} \rightarrow \infty} d_S^{(2+1)}(\sigma) = 3$, $\lim_{\kappa\sqrt{\sigma} \rightarrow 0} d_S^{(2+1)}(\sigma) = 2$, respectively. In 1+1 dimensions (\ref{eq:4.7b}) leads to
\begin{align}\label{eq:4.11}
d_S^{(1+1)}(\sigma) = 1 + 2 \kappa^2 \sigma \left(\frac{1}{\sqrt{\pi}\, \kappa \sqrt{\sigma}} \frac{e^{-\kappa^2 \sigma}}{1 - {\rm erf}(\kappa \sqrt{\sigma})} - 1\right)\,,
\end{align}
with the IR and UV limits $\lim_{\kappa\sqrt{\sigma} \rightarrow \infty} d_S^{(1+1)}(\sigma) = 2$, $\lim_{\kappa\sqrt{\sigma} \rightarrow 0} d_S^{(1+1)}(\sigma) = 1$, respectively.

\subsection{Bicrossproduct Laplacian} \label{sec:4.2}

Another possible Laplacian on $\kappa$-Minkowski space, called the bicrossproduct Laplacian, corresponds to the simplest mass Casimir of $\kappa$-Poincar\'{e} algebra \cite{Lukierski:1992ny} (let us stress that any function of this Casimir that has the correct classical limit is also a Casimir). More precisely, it is the Euclidean version of the momentum space representation of the Casimir and has the form
\begin{align}\label{eq:4.12}
{\cal L}_{\text{cp}}(P_0,\{P_a\}) = 4\kappa^2 \sinh^2\left(\tfrac{1}{2\kappa} P_0\right) + e^{P_0/\kappa} P_aP^a\,.
\end{align}
In classical coordinates it becomes
\begin{align}\label{eq:4.12a}
{\cal L}_{\text{cp}}(p_0, \{p_a\}) = 2\kappa \left(\sqrt{p_0^2 + p_ap^a + \kappa^2} - \kappa\right).
\end{align}
It turns out that the corresponding return probability in 3+1 dimensions is given by a simple rational function
\begin{align}\label{eq:4.13}
{\cal P}_{(3+1)}(\sigma) = \pi^2 \frac{2\kappa^2 \sigma + 1}{2\kappa^2 \sigma^3}\,,
\end{align}
while in 2+1 dimensions we obtain (\cite{Arzano:2014de} in this case had only numerical results)
\begin{align}\label{eq:4.13a}
{\cal P}_{(2+1)}(\sigma) = \frac{4\pi \kappa}{\sigma}\, e^{2\kappa^2 \sigma} K_1(2\kappa^2 \sigma)\,,
\end{align}
where $K_\alpha(\cdot)$ is a modified Bessel function of the second kind. Finally, in 1+1 dimensions we have just
\begin{align}\label{eq:4.13b}
{\cal P}_{(1+1)}(\sigma) = \frac{\pi}{\sigma}.
\end{align}
Observe that in the latter case there is no dependence on the parameter $\kappa$.

The dimension spectra can again be obtained directly from the expansions around $\sigma = 0$ of the exact formulae (\ref{eq:4.13}--\ref{eq:4.13a}) (and the trivial (\ref{eq:4.13b})):
\begin{align}\label{eq:4.15}
{\cal P}_{(3+1)}(\sigma) &= \frac{\pi^2}{2\kappa^2 \sigma^3} + \frac{\pi^2}{\sigma^2}\,, \nonumber\\
{\cal P}_{(2+1)}(\sigma) &= \frac{2\pi}{\kappa \sigma^2} + \frac{4\kappa \pi}{\sigma} + 4\kappa^3 \pi \log\sigma + 2\kappa^3 \pi \left(1 + 2\gamma + 4\log\kappa\right) + {\cal O}(\sigma)\,.
\end{align}
Consequently, we have
\begin{align}\label{eq:4.16}
{\rm Sd}_{(3+1)} &= \{3,2\}\,, & \ord \Sd = 1\,,\nonumber\\
{\rm Sd}_{(2+1)} &= 2-\mathbb{N} = \{2,1,0,-1,-2,\ldots\}\,, & \ord \Sd = 2\,, \nonumber\\
{\rm Sd}_{(1+1)} &= \{1\}\,, & \ord \Sd = 1\,.
\end{align}

If we use the standard formula (\ref{eq:3.5}), (\ref{eq:4.13}) leads to the spectral dimension
\begin{align}\label{eq:4.14}
d_S^{(3+1)}(\sigma) = 6 - \frac{4\kappa^2 \sigma}{2\kappa^2 \sigma + 1}\,,
\end{align}
with $\lim_{\kappa\sqrt{\sigma} \rightarrow \infty} d_S^{(3+1)}(\sigma) = 4$ and $\lim_{\kappa\sqrt{\sigma} \rightarrow 0} d_S^{(3+1)}(\sigma) = 6$; and (\ref{eq:4.13a}) leads to
\begin{align}\label{eq:4.14a}
d_S^{(2+1)}(\sigma) = 4 - 4\kappa^2 \sigma \left(1 - \frac{K_0(2\kappa^2 \sigma)}{K_1(2\kappa^2 \sigma)}\right)\,,
\end{align}
with $\lim_{\kappa\sqrt{\sigma} \rightarrow \infty} d_S^{(2+1)}(\sigma) = 3$ and $\lim_{\kappa\sqrt{\sigma} \rightarrow 0} d_S^{(2+1)}(\sigma) = 4$. In both cases we observe the dimension growing at small scales above the classical value (see, however, Subsec.~\ref{sec:4.4}), which -- from the perspective of a random walk process (\ref{eq:3.1}) -- is the pattern of superdiffusion. Finally, the case of 1+1 dimensions is trivial  -- with no dimensional flow.

\subsection{Relative-locality Laplacian} \label{sec:4.3}

The last Laplacian that is of our interest has been proposed in the framework of relative locality. This relative-locality Laplacian is determined by the square of the geodesic distance from the origin in momentum space \cite{Gubitosi:2013re}, which has the same form for both the Lorentzian and Euclidean metric signature \cite{Arzano:2014de}, namely $d^2(p_0,p_a) = \kappa^2 \mathrm{arccosh}^2 (p_{-1}/\kappa)$. In bicrossproduct coordinates it becomes
\begin{align}\label{eq:4.17}
{\cal L}_{\text{rl}}(P_0,\{P_a\}) = -\kappa^2 \mathrm{arccosh}^2 \left(\cosh\left(\tfrac{1}{\kappa} P_0\right) + \frac{1}{2\kappa^2} e^{P_0/\kappa} P_aP^a\right)\,,
\end{align}
while in classical coordinates,
\begin{align}\label{RL}
{\cal L}_{\text{rl}}(p_0,\{p_a\}) = -\kappa^2 \mathrm{arccosh}^2 \left(\tfrac{1}{\kappa} \sqrt{p_0^2 + p_a p^a + \kappa^2}\right)\,.
\end{align}
The return probability in 3+1 dimensions is now given by
\begin{align}\label{eq:4.18}
{\cal P}_{(3+1)}(\sigma) = \frac{\pi^{5/2} \kappa^3}{4\sqrt{\sigma}} e^{1/(4\kappa^2 \sigma)} \left(e^{2/(\kappa^2 \sigma)} {\rm erf}\left(\frac{3}{2\kappa \sqrt{\sigma}}\right) - 3\, {\rm erf}\left(\frac{1}{2\kappa \sqrt{\sigma}}\right)\right)\,,
\end{align}
in 2+1 dimensions by
\begin{align}\label{eq:4.18a}
{\cal P}_{(2+1)}(\sigma) = \frac{\pi^{3/2} \kappa^2}{\sqrt{\sigma}} \left(e^{1/(\kappa^2 \sigma)} - 1\right)
\end{align}
and in 1+1 dimensions by
\begin{align}\label{eq:4.18b}
{\cal P}_{(1+1)}(\sigma) = \frac{\pi^{3/2} \kappa}{\sqrt{\sigma}}\, e^{1/(4\kappa^2 \sigma)} {\rm erf}\left(\frac{1}{2\kappa \sqrt{\sigma}}\right)
\end{align}
(\cite{Arzano:2014de} had only numerical results for the Laplacian ${\cal L}_{\text{rl}}$). 

In this case, neither of the heat traces (\ref{eq:4.18}--\ref{eq:4.18b}) can be expanded in a series of the form \eqref{HeatGen}. This is because of the factor $e^{1/\sigma}$, which yields an essential singularity at $\sigma = 0$. Consequently, the dimension spectrum of the relative-locality Laplacian does not exist in any of the three considered topological dimensions. 

On the other hand, given the exact formulae (\ref{eq:4.18}--\ref{eq:4.18b}), we can compute the corresponding spectral dimensions explicitly:
\begin{align}\label{eq:4.19}
d_S^{(3+1)}(\sigma) = 1 + \frac{3}{2\kappa^2 \sigma} \frac{{\rm erf}\left(\frac{1}{2\kappa \sqrt{\sigma}}\right) - 3\, e^{2/(\kappa^2 \sigma)} {\rm erf}\left(\frac{3}{2\kappa \sqrt{\sigma}}\right)}{3\, {\rm erf}\left(\frac{1}{2\kappa \sqrt{\sigma}}\right) - e^{2/(\kappa^2 \sigma)} {\rm erf}\left(\frac{3}{2\kappa \sqrt{\sigma}}\right)}\,,
\end{align}
with $\lim_{\kappa\sqrt{\sigma} \rightarrow \infty} = 4$, $\lim_{\kappa\sqrt{\sigma} \rightarrow 0} = \infty$; similarly, in 2+1 dimensions
\begin{align}\label{eq:4.19a}
d_S^{(2+1)}(\sigma) = 1 + \frac{2}{\kappa^2 \sigma} \frac{1}{1 - e^{-1/(\kappa^2 \sigma)}}\,,
\end{align}
with $\lim_{\kappa\sqrt{\sigma} \rightarrow \infty} = 3$, $\lim_{\kappa\sqrt{\sigma} \rightarrow 0} = \infty$; and in 1+1 dimensions
\begin{align}\label{eq:4.19b}
d_S^{(1+1)}(\sigma) = 1 + \frac{1}{2\kappa^2 \sigma} \left(1 + \frac{2\kappa \sqrt{\sigma}\, e^{-1/(4\kappa^2 \sigma)}}{\sqrt{\pi}\, {\rm erf}\left(\frac{1}{2\kappa \sqrt{\sigma}}\right)}\right)\,,
\end{align}
with $\lim_{\kappa\sqrt{\sigma} \rightarrow \infty} = 2$, $\lim_{\kappa\sqrt{\sigma} \rightarrow 0} = \infty$. The UV divergence of $d_S(\sigma)$ might be considered problematic, but such behaviour of the dimensionality (sometimes also for the Hausforff dimension) has been encountered in other models of quantum spacetime and at least in commutative geometry seems to be a consequence of the extremely high connectivity between points of space \cite{Mandrysz:2019us}.

\subsection{Comparison} \label{sec:4.4}

For the sake of comparison, let us first recall \eqref{eq:3.4} that the standard return probability on $\mathbb{R}^n$ reads:
\begin{align}
\P_{\text{class}}(\sigma) = (4 \pi \sigma)^{-n/2}.
\end{align} 
Hence, the dimension spectrum consists of a single element $\{n/2\}$ and is of order 1. The spectral dimension is a constant function equal to $n$.

The first observation is that for the bicovariant Laplacian the dimension spectra contain multiple elements. By analogy with the Riemannian geometry, one could interpret this as a signature of some sort of curved geometry of $\kappa$-Minkowski space. Furthermore, in the 2+1 dimensional case the dimension spectrum is of order 2. From the perspective of (pseudo)differential geometry, this means that the heat trace expansion involves non-local coefficients. The situation is analogous for the bicrossproduct Laplacian, apart from the 1+1 dimensional case, for which the dimension spectrum coincides with the classical one. The relative locality Laplacian does not have a dimension spectrum at all, which suggests that the corresponding geometry is infinite dimensional.

In stark contrast with the quantum spheres, none of the studied Laplacians on $\kappa$-Minkowski space exhibits complex numbers outside of the real axis in its dimension spectrum. Consequently, there are no oscillations in the corresponding spectral dimensions. The latter is also true for the relative locality Laplacian.

Let us now inspect the spectral dimensions closer.

\begin{figure}[h]
\centering
\includegraphics[width=0.45\textwidth]{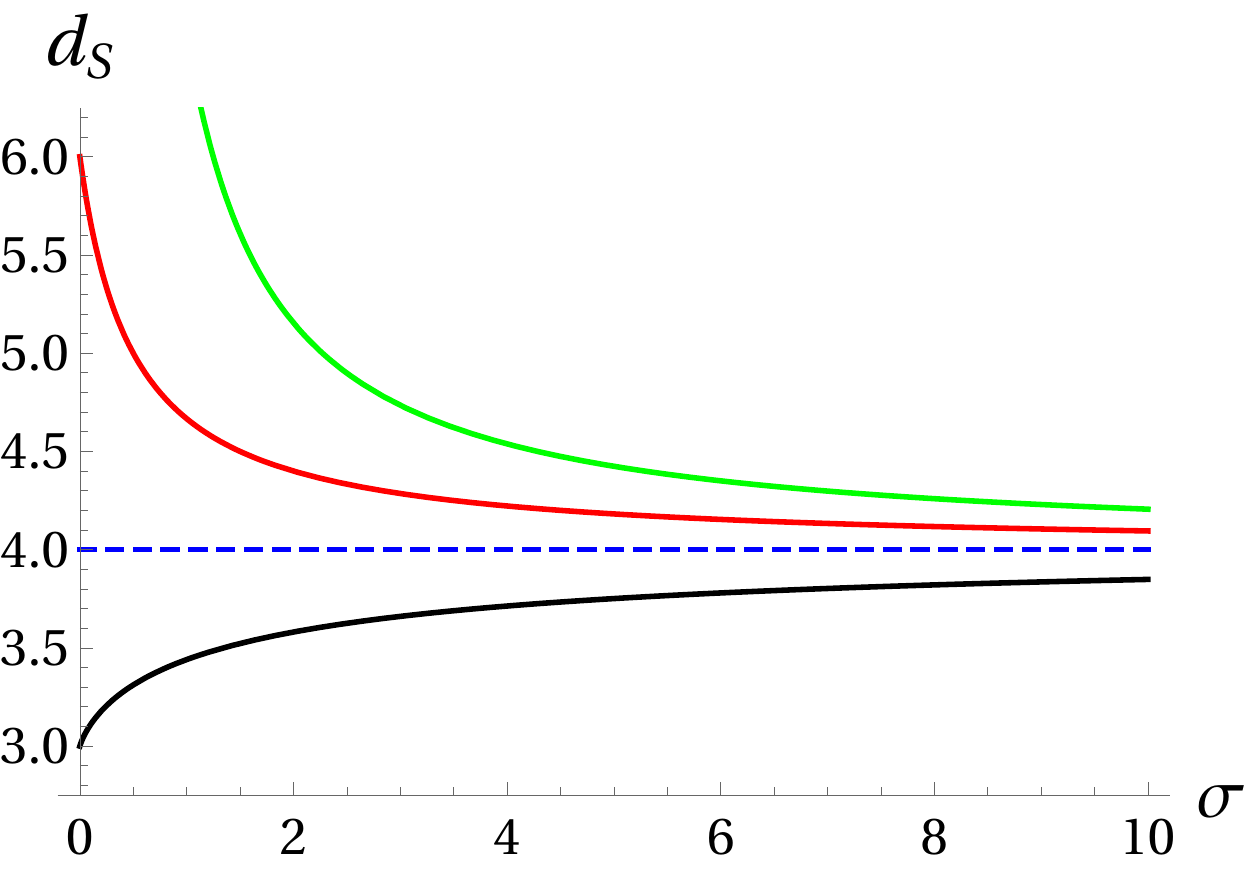}
\hspace{0.05\textwidth}
\includegraphics[width=0.45\textwidth]{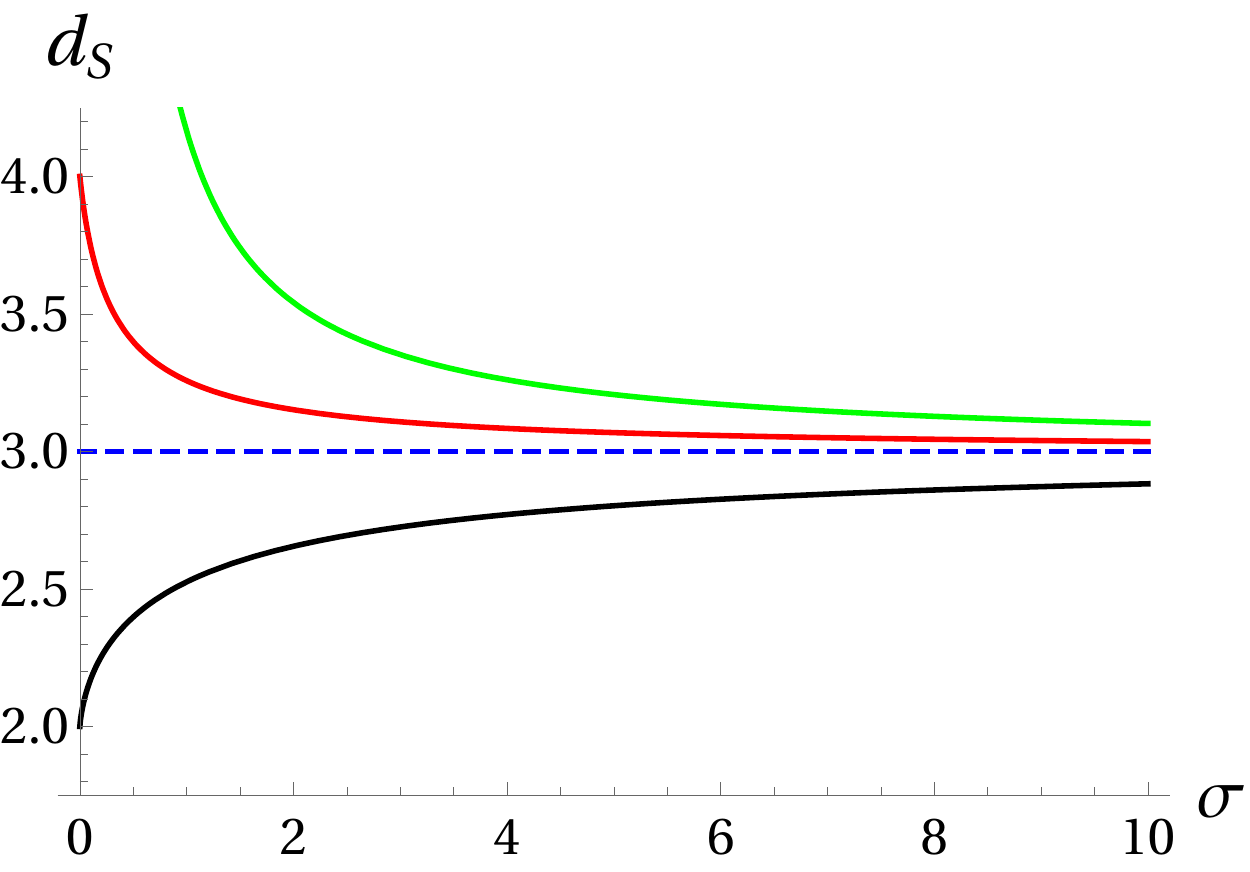}
\caption{\label{dS_spheres}Spectral dimension $d_S(\sigma)$ in 3+1 dimensions (left) and in 2+1 dimensions (right) for the Laplacians ${\cal L}_{\text{cv}}$ (bottom curve), ${\cal L}_{\text{cp}}$ (middle curve) and ${\cal L}_{\text{rl}}$ (top curve) (we set $\kappa = 1$).}
\label{fig:03}
\end{figure}

The profiles of $d_S(\sigma)$ for different Laplacians in 3+1 and 2+1 topological dimensions are compared in Fig.~\ref{fig:03} (the situation in 1+1 topological dimensions is qualitatively the same apart from the case of ${\cal L}_{\text{cp}}$, for which $d_S(\sigma) = 2$ --- cf. Subsec.~\ref{sec:4.2}). All curves in both plots exhibit the strong discrepancy in the UV, while they converge to the same IR limit, equal to the topological dimension. Unless we have some extra reason to claim that only one Laplacian is physically correct, the spectral dimension seems to be an ambiguous characteristic of $\kappa$-Minkowski space. Some further comments about the relations between specific UV limits visible in Fig.~\ref{fig:03} and various results obtained within the quantum gravity research, which could single out one of the Laplacians, can be found in \cite{Arzano:2014de}. We also note that a recent analysis \cite{Arzano:2017ne} of a static potential between two sources on 3+1-dimensional $\kappa$-Minkowski space brings evidence that the physical dimension in the UV is equal to 3, in agreement with the result for the bicovariant Laplacian.

Let us note, however, that the spectral dimensions $d_S(\sigma)$ in the previous subsections were calculated under an implicit assumption that all three operators ${\cal L}$ are of order 2 --- recall the discussion around Eq. \eqref{dSm}. This is based on the fact that these operators are deformations of the classical Laplacian, which is a second order differential operator. In the case of the quantum spheres, we could provide an external argument for the order of quantum Laplacians, which is based on a rigorous first-order condition in the theory of spectral triples. In the case of $\kappa$-Minkowski space such an argument is not available, because the considered Laplacians do not originate from a Dirac operator.

Observe that in classical coordinates the bicovariant Laplacian acquires the same form \eqref{eq:4.6a} as the standard Laplacian on $\sR^{n+1}$. This justifies the assumption that its order equals 2. On the other hand, the bicrossproduct Laplacian in classical coordinates \eqref{eq:4.12a} looks as if was of order 1. Even more curiously, the relative-locality Laplacian in classical coordinates \eqref{RL} seems to be of order 0. Indeed, its leading behaviour for large values of $\vert p \vert \vc \sqrt{p_0^2 + p_a p^a}$ is $\log \vert p \vert$, which grows slower than any power of $\vert p \vert$.

The problem is that in the quantum gravity models one usually considers the quantum spacetime to be a certain deformation of the classical one (more accurately, it is the classical spacetime that is an approximation of the quantum one), with the deformation controlled by some parameter(s) related to the Planck length or mass. In particular, it is expected that a generalised Laplacian becomes the standard one in the classical limit, when the deformation vanishes, i.e. we actually have a parametrized family of operators, valid at different scales. In this context, Formula (\ref{eq:3.5}) allows one to track a deviation from the IR value of spacetime dimension. Using (\ref{dSm}) with $\eta \neq 2$ would prevent us from recovering the correct IR limit, unless we allow the order of the operator to depend on the deformation parameter.

If we assumed the order of the bicrossproduct Laplacian to be a continuous function of $\kappa$, such that $\lim_{\kappa \to 0} \eta(\kappa) = 1$ and $\lim_{\kappa \to \infty} \eta(\kappa) = 2$, then we would obtain the dimension $n$ in the UV and $n+1$ in the IR, as in the bicovariant case. In the same vein, we could set the order of the relative-locality Laplacian to be a function of $\kappa$,  satisfying $\lim_{\kappa \to \infty} \eta(\kappa) = 2$. The situation in the UV limit in this case is more ambiguous since the dimension diverges, while the order should tend to zero and their ratio could yield any number. In particular, the behaviour of $\eta(\kappa)$ close to $\kappa = 0$ might be such that $\lim_{\kappa \to 0} 2/\eta(\kappa)\, d_S(\sigma) = n$, as for other Laplacians. Taking $d_S(\sigma) = d_S(\kappa^2 \sigma)$ given by (\ref{eq:4.19}-\ref{eq:4.19b}) and considering the Ansatz $\eta(\kappa) \propto \kappa^2$, we find that the desired limit requires $\eta(\kappa) \sim \tfrac{4}{n} \kappa^2$ for $\kappa \to 0$. Such a trick allows us to remove the discrepancy between the UV behaviour of the spectral dimensions of all three Laplacians, while maintaining the correct classical limit. It also agrees with the viewpoint on the operator orders suggested by the formulae in classical coordinates. 

Nevertheless, let us stress that the deformation-parameter-dependent order is only a tentative hypothesis, introduced by us here. Furthermore, even if this hypothesis was rigorously implemented and erased differences between spectral dimensions for the three Laplacians, the corresponding dimension spectra would still remain irreconcilably different. For example, the dimension spectrum in the case of 3+1 topological dimensions, depending on the Laplacian, is infinite, has only two elements or does not exist.

\section{\label{sec:discussion}Summary}

Our analysis shows that the UV behaviour of the spectral dimensions of quantum spacetimes is fairly complex and not limited to a monotonic flow. To understand it we have adopted from noncommutative geometry \`a la Connes a rigorous notion of a dimension spectrum. The latter characterises the UV behaviour of a heat trace, from which the spectral dimension is deduced. The relationship between the spectral dimension and the dimension spectrum of a given Laplacian (more generally, an operator of order $\eta$) is captured by the following dictionary:

\begin{itemize}

	\item The dimension spectrum consisting of more than one element implies a non-constant behaviour of the spectral dimension and suggests a non-trivial geometry.

	\item If the UV limit $\sigma \to 0$ is finite, then $d_S(0) = \eta \, d_{\Sd}$, where $d_{\Sd}$ is largest real number in the dimension spectrum \eqref{dSd}. Note that, in general, $d_{\Sd}$ need not be a natural number. This happens routinely in fractal geometry, where $d_{\Sd}$ recovers the Hausdorff dimension  \cite{ConnesGarden,Lapidus,Kellendonk,CGIS,Christensen1}.
	
	\item The order of the dimension spectrum determines the leading behaviour of the spectral dimension for small $\sigma$. If $\ord\Sd > 1$, then this behaviour is logarithmic. More concretely, if $\P(\sigma) \sim (\log \sigma)^p \sigma^{-r}$ as $\sigma$ tends to 0, with $p = \ord \Sd - 1 \in \N$ and $r \geq 0$, then $d_S(\sigma) \sim 2r - 2p / (\log \sigma)$. This happens on the quantum sphere with $p=2$ and $r=0$. On the other hand, a sub-leading logarithmic behaviour $\P(\sigma) \sim \alpha \sigma^{-r} + \beta \sigma^{-r-1} (\log \sigma)^p$ translates to $d_S(\sigma) \sim 2 (r-1) + 2\alpha / (\alpha + \beta \sigma (\log \sigma)^p)$. This is exemplified on the 2+1-dimensional $\kappa$-Minkowski spacetime, with $p = 1$ and $r = 1$ or $r = 2$ for the bicovariant and the bicrossproduct Laplacians, respectively.
	
	\item The presence of non-real numbers in $\Sd$ signifies the presence of log-periodic oscillations in the UV behaviour of the spectral dimension.
	
\end{itemize}

With the help of the dimension spectrum we have detected the log-periodic oscillations of the spectral dimension for the quantum sphere. These were overlooked in \cite{Benedetti:2009fe}, because of their very small amplitude for the deformation parameter close to the classical value. It is remarkable that the complex dimensions occur here in a somewhat unexpected setting, where no fractal properties or discrete scale invariance were a priori imposed.

In contrast, such oscillations are present on $\kappa$-Minkowski spacetime for none of the 3 considered Laplacians, at least in the cases of 2, 3 and 4 topological dimensions that we studied. One might relate this to the fact the quantum sphere is compact, whereas $\kappa$-Minkowski spacetime is not and interpret the oscillations as a peculiar IR/UV mixing effect. On the other hand, from a more conservative standpoint, one could simply say that the quantum sphere is ``more quantum''. In either case, it seems worth looking for the complex dimensions in other models of quantum spacetime.

Finally, we would like to stress that the spectral dimension and dimension spectrum of a given quantum spacetime strongly depend on the chosen Laplacian. On the mathematical side, this is simply a consequence of their definitions, arising from the trace of the heat operator, which is constructed from a given Laplacian. As we discussed in the last subsection, the discrepancy between the UV behaviour of the spectral dimensions of $\kappa$-Minkowski space with different Laplacians can perhaps be removed using the concept of the deformation-dependent order of the Laplacian but even then, differences will survive in the dimension spectra. On the other hand, while the dimension spectra of the quantum sphere with the spinorial and scalar Laplacians are identical, their spectral dimensions do not overlap and actually behave in the opposite ways in the IR. Thus, the spectral dimension and dimension spectrum provide complementary information about a quantum spacetime structure.

In terms of physics, we believe that it is of key importance to understand which aspects of dimensionality are genuine to the given quantum geometry and which are just mathematical artefacts of the chosen Laplacian. Our analysis suggests that such universal properties are the log-periodic oscillations for the quantum sphere, as well as the lack of these for the $\kappa$-Minkowski spacetime, because these features persist for all of the studied Laplacians. The same could be said about the appearance of third order poles in the dimension spectra on quantum sphere. As pointed out, this could be taken as an indication of some sort of `singular' or `non-smooth' character of this quantum spacetime. On the other hand, the dimension spectra of $\kappa$-Minkowski spacetime fit within the ones obtained in classical (pseudo)differential geometry. Interestingly enough, the second order poles, and hence the $\log \sigma$ terms in the heat trace, occur only in 2+1 topological dimensions, but for both bicovariant and bicrossproduct Laplacians. This suggests that the geometry of 2+1 dimensional $\kappa$-Minkowski spacetime departs more strongly from the classical case. However, further studies --- possibly connected with the expected non-local nature of some coefficients of the heat trace expansion --- would be needed to confirm this conclusion.

\appendix

\section{\texorpdfstring{The spectral zeta function of the operator $\Delta_q\scl$}{The spectral zeta function of the quantum scalar Laplacian}}\label{app}

In this appendix we provide an explicit meromorphic extension of the spectral zeta function associated with the operator $\Delta_q\scl$ defined by Formula \eqref{Delta_scal}. Before we start we need to take care of the zero mode, as for any $s \in \sC$ the operator $(\Delta_q\scl)^{-s}$ is not trace-class on the full Hilbert space $\H_q'$. We thus set $\h \vc (\ker \Delta_q\scl)^{\perp}$ and compute the zeta function without the zero mode. Note that after such a truncation, it is still possible to use the inverse Mellin transform technique for the computation of the corresponding heat trace (see \cite{HeatEZ} for the general method). Indeed, we have
\begin{align*}
\Pscl (\sigma) & = \Tr_{\H_q'} e^{- \sigma \Delta_q\scl} = 1 + \Tr_{\h} e^{- \sigma \Delta_q\scl}
\end{align*}
and simply use formulae (\ref{Mell}--\ref{poles}) with $\h$ at the place of $\H_q'$. 

Now, for $\Re(s) > 0$ we have
\begin{align*}
\zeta_{\Delta_q\scl}(s) &= \Tr_{\h}\, (\Delta_q\scl)^{-s} = \sum_{j=1}^{\infty} \sum_{m=-j}^{j} \bra{j,m} (\Delta_q\scl)^{-s} \ket{j,m} \\
&  = u(\sqrt{q})^{-s} \sum_{j=1}^{\infty} (2j+1) q^{s/2} \left( q^{-j} - 1 - q + q^{j+1} \right)^{-s} \\
& = \big( u(\sqrt{q}) q^{-3/2} \big)^{-s} \sum_{k=0}^{\infty} (2k+3) \left( q^{-k} - q - q^2 + q^{k+3} \right)^{-s} \\
& = \big( u(\sqrt{q}) q^{-3/2} \big)^{-s} \sum_{k=0}^{\infty} (2k+3) q^{ks} \left( 1 - q^{k+1} \right)^{-s} \left( 1 - q^{k+2} \right)^{-s}.
\end{align*}

In order to construct a meromorphic extension of $\zeta_{\Delta_q\scl}$ to the entire complex plane we use the standard binomial expansion formula
\begin{align}\label{binom}
(1-x)^{-s} = \sum_{n = 0}^{\infty} \binom{s+n-1}{n} x^n\,
\end{align}
valid for any complex number $s$ and any $x \in \sC$ with $\vert x \vert < 1$. The coefficients $\binom{s+n-1}{n} = \tfrac{\Gamma(s+n)}{n! \, \Gamma(s)}$ are polynomials in $s$ of order $n$.

With the help of formula \eqref{binom} we rewrite the zeta function as follows:
\begin{align*}
\zeta_{\Delta_q\scl}(s) &= \big( u(\sqrt{q}) q^{-3/2} \big)^{-s} \sum_{k=0}^{\infty} (2k+3) \sum_{\ell = 0}^{\infty} \sum_{m = 0}^{\infty} \binom{s+\ell-1}{\ell} \binom{s+m-1}{m} q^{(k+1)\ell} q^{(k+2)m} q^{ks}.
\end{align*}
For $\Re(s) > 0$ all of the series are absolutely convergent and we are free to change the order of summation and compute the sum over $k$. We thus have
\begin{align*}
\zeta_{\Delta_q\scl}(s) &= \big( u(\sqrt{q}) q^{-3/2} \big)^{-s} \sum_{\ell = 0}^{\infty} \sum_{m = 0}^{\infty} \binom{s+\ell-1}{\ell} \binom{s+m-1}{m} \frac{q^{\ell + 2m} (3- q^{\ell + m +s})}{(1-q^{\ell+m+s})^2} \\
&= \big( u(\sqrt{q}) q^{-3/2} \big)^{-s} \sum_{n = 0}^{\infty} \sum_{m = 0}^{n} \binom{s+n-m-1}{n-m} \binom{s+m-1}{m} \frac{q^{n+m} (3- q^{n +s})}{(1-q^{n+s})^2} \\
&= \big( u(\sqrt{q}) q^{-3/2} \big)^{-s} \sum_{n = 0}^{\infty} \frac{q^{n} (3- q^{n +s})}{(1-q^{n+s})^2}   \binom{s+n-1}{n} {}_2F_1(-n,s;-n+s+1;q),
\end{align*}
with a hypergeometric function $_2F_1$.

The last series over $n$ can easily be shown (cf. \cite[Proposition 3.2]{PodlesSA}) to be absolutely convergent for any complex $s$ outside of the discrete set $\tfrac{\pi i}{\log q} \sZ - \sN$. We have thus obtained a meromorphic extension of $\zeta_{\Delta_q\scl}$ to the entire complex plane. It has isolated double poles precisely in the set $\Sd \Delta_q\scl = \Sd \Delta_q\spin = \tfrac{\pi i}{\log q} \sZ - \sN$. The poles of the Gamma function contribute additional poles at $-n$, $n \in \sN$, hence the order of the dimension spectrum is 3.

Formula \eqref{z1} for the meromorphic extension of the function $\zeta_{\Delta_q\spin}$ is proved along the same lines with the help of the identity \eqref{binom}. The two zeta functions have the same meromorphic structure, though $\zeta_{\Delta_q\scl}$ has a more involved form of the coefficients.

\begin{acknowledgments}
The work of ME was supported by the National Science Centre in Poland under the research grant Sonatina (2017/24/C/ST2/00322). Publication supported by the John Templeton Foundation Grant ,,Conceptual Problems in Unification Theories'' (No. 60671).
\end{acknowledgments}

\bibliography{heat_dim}
\end{document}